\documentclass[nofootinbib,11pt]{revtex4-1}
\usepackage{amsmath,amssymb,pgf,pgfarrows,pgfnodes,float,appendix, hyperref,slashed,breakurl,graphicx}
\usepackage{graphicx}
\usepackage{subfigure}
\usepackage[margin=0.9in]{geometry}
\usepackage{ulem}
\usepackage{placeins}

\newcommand{\SU}{\text{SU}}

\usepackage{color, colortbl}
\definecolor{Gray}{gray}{0.8}
\definecolor{GrayLight}{gray}{0.4}
\definecolor{Darkgreen}{RGB}{30,120,30}
\definecolor{Orange}{rgb}{1,0.38,0.11}

\newcommand{\beq}{\begin{equation}}
\newcommand{\eeq}{\end{equation}}
\newcommand{\bea}{\begin{eqnarray}}
\newcommand{\eea}{\end{eqnarray}}

\usepackage{tikz}
\usetikzlibrary{"arrows", "automata", "backgrounds", "calendar", "chains", "matrix", "mindmap", "patterns", "petri", "shadows", "shapes.geometric", "shapes.misc", "spy", "trees"}
\usetikzlibrary{arrows,shapes}
\usetikzlibrary{trees}
\usetikzlibrary{matrix,arrows} 				
\usetikzlibrary{positioning}				
\usetikzlibrary{calc,through}				
\usetikzlibrary{decorations.pathreplacing}  
\usepackage{pgffor}							

\usetikzlibrary{decorations.pathmorphing}	
\usetikzlibrary{decorations.markings}
\tikzset{
    vector/.style={decorate, decoration={snake}, draw},
	provector/.style={decorate, decoration={snake,amplitude=2.5pt}, draw},
	antivector/.style={decorate, decoration={snake,amplitude=-2.5pt}, draw},
    fermion/.style={draw=black, postaction={decorate},
        decoration={markings,mark=at position .55 with {\arrow[draw=black]{>}}}},
    fermioncyan/.style={draw=black, postaction={decorate},
        decoration={markings,mark=at position .55 with {\arrow[draw=cyan]{<}}}},
    fermiondif/.style={draw=black, postaction={decorate},
        decoration={markings,mark=at position .7 with {\arrow[draw=black]{<}}}},
    fermionend/.style={draw=black, postaction={decorate},
        decoration={markings,mark=at position 1 with {\arrow[draw=black]{>}}}},
    fermionuchannel2/.style={draw=black, postaction={decorate},
        decoration={markings,mark=at position .4 with {\arrow[draw=black]{>}}}},
    scalardif/.style={dashed,draw=black, postaction={decorate},
        decoration={markings,mark=at position .7 with {\arrow[draw=black]{>}}}},
    scalarend/.style={dashed,draw=black, postaction={decorate},
        decoration={markings,mark=at position 1 with {\arrow[draw=black]{>}}}},
    fermionbar/.style={draw=black, postaction={decorate},
        decoration={markings,mark=at position .55 with {\arrow[draw=black]{<}}}},
    fermionnoarrow/.style={draw=black},
    gluon/.style={decorate, draw=black,
        decoration={coil,amplitude=4pt, segment length=5pt}},
    scalar/.style={dashed,draw=black, postaction={decorate},
        decoration={markings,mark=at position .55 with {\arrow[draw=black]{>}}}},
    scalarcyan/.style={dashed,draw=black, postaction={decorate},
        decoration={markings,mark=at position .55 with {\arrow[draw=cyan]{>}}}},
    scalaruchannel1/.style={dashed,draw=black, postaction={decorate},
        decoration={markings,mark=at position .7 with {\arrow[draw=black]{>}}}},
                  scalaruchannel2/.style={dashed,draw=black, postaction={decorate},
        decoration={markings,mark=at position .4 with {\arrow[draw=black]{>}}}},
    scalarbar/.style={dashed,draw=black, postaction={decorate},
        decoration={markings,mark=at position .55 with {\arrow[draw=black]{<}}}},
    scalarnoarrow/.style={dashed,draw=black},
    electron/.style={draw=black, postaction={decorate},
        decoration={markings,mark=at position .55 with {\arrow[draw=black]{>}}}},
	bigvector/.style={decorate, decoration={snake,amplitude=4pt}, draw},
}

\tikzstyle{block} = [draw, rectangle, 
    minimum height=3em, minimum width=6em]

\usepackage{xparse}
\NewDocumentCommand\semiloop{O{black}mmmO{}O{above}}
{%
\draw[#1] let \p1 = ($(#3)-(#2)$) in (#3) arc (#4:({#4+180}):({0.5*veclen(\x1,\y1)})node[midway, #6] {#5};)
}

\tikzset{
    cross/.pic = {
    \draw[rotate = 45] (-#1,0) -- (#1,0);
    \draw[rotate = 45] (0,-#1) -- (0, #1);
    }
}

\NewDocumentCommand\semilooptwo{O{black}mmmO{}O{above}}
{%
\draw[#1] let \p1 = ($(#3)-(#2)$) in (#3) arc (#4:({#4-180}):({0.5*veclen(\x1,\y1)})node[midway, #6] {#5};)
}

\tikzstyle{block} = [draw, rectangle, 
    minimum height=3em, minimum width=6em]

\begin{document}

\hspace{5.2in} \mbox{CALT-TH/2021-044}

\title{Dark Unification: a UV-complete Theory of Asymmetric Dark Matter}
\author{Clara Murgui}
\affiliation{Walter Burke Institute for Theoretical Physics, California Institute of Technology, Pasadena, CA USA }
\author{Kathryn M. Zurek}
\affiliation{Walter Burke Institute for Theoretical Physics, California Institute of Technology, Pasadena, CA USA }

\begin{abstract}

Motivated by the observed ratio of dark matter to baryon mass densities, $\rho_D/\rho_B \simeq 5$, we propose a theory of dark-color unification.  In this theory, the dark to visible baryon masses are fixed by the ratio of dark to visible confinement scales, which are determined to be nearby in mass through the unification of the dark and visible gauge theories at a high scale.  Together with a mechanism for darko-baryo-genesis, which arises naturally from the grand unification sector, the mass densities of the two sectors must be nearby, explaining the observed mass density of dark matter.  We focus on the simplest possible example of such a theory, where Standard Model color $\SU(3)_C$ is unified with dark color $\SU(2)_D$ into $\SU(5)$ at an intermediate scale of around $10^8-10^9$~GeV.  The dark baryon consists of two dark quarks in an isotriplet configuration.  There are a range of important cosmological, astrophysical and collider signatures to explore, including dark matter self-interactions, early matter domination from the dark hadrons, gravitational wave signatures from the hidden sector phase transition, contributions to flavor observables, as well as Hidden Valley-like signatures at colliders. 

\end{abstract}

\maketitle
\tableofcontents

\section{Introduction}

In theories where the dark matter (DM) density is related to the baryon asymmetry, the relic abundance of dark matter is fixed by its particle-anti-particle asymmetry, similar to the baryon asymmetry. The chemical potential of the dark matter and baryons is related, such that the number densities are related by an ${\cal O}(1)$ number, $c$:
\beq
n_B = c \, n_D,
\eeq
thus predicting a mass ratio
\beq
\frac{m_\text{DM}}{m_p} = c \, \frac{\rho_D}{\rho_B} \approx 5 \, c. \label{eq:energy}
\eeq
This sets the natural mass for the dark matter, dependent on $c$.  

Most theories with a DM asymmetry do not purport to explain why the dark sector mass gap is close to that of Standard Model (SM) QCD.  The first efforts to relate the DM density to the baryon density mostly relied on physics at the electroweak scale~\cite{Nussinov:1985xr,Barr:1990ca,Kaplan:1991ah,Gudnason:2006yj,Khlopov:2007ic}, with a Boltzmann suppression factor to generate the needed hierarchy of about two orders of magnitude between the electroweak and the QCD scale. Attempts to justify the similar scale for QCD and the dark matter mass have considered a mirror QCD, which is entwined with the SM QCD by an exact~\cite{Foot:2003jt,Foot:2004pq} or spontaneously broken~\cite{Lonsdale:2014wwa,Lonsdale:2014yua,Lonsdale:2018xwd} ${\cal Z}_2$ mirror symmetry.  The reliance on the Standard Model and related dynamics for the physics of the DM generically places strong constraints.

Within the context of UV complete theories such as string theory and Grand Unification, however, the dark matter sector is generically not parasitic on the Standard Model dynamics.  The hidden sector may contain matter content that is strongly or weakly coupled, or both.  Its complexity and mass gaps can be independent of the Standard Model, or coupled to it through messenger states that mediate dark sector - Standard Model interactions.  The dark sector mass gap can be quite low while still retaining complexity in its dynamics, and couplings to the SM through higher dimension operators.  This idea was at the heart of Hidden Valley models~\cite{Strassler:2006im}, and in the proposal of Asymmetric Dark Matter (ADM)~\cite{Kaplan:2009ag} as the low energy incarnation of a high scale messenger sector that shares dark and visible baryon number but decouples to separately freeze-in the asymmetries.  

These hidden sector theories, by virtue of the fact that the dynamics of the two sectors are not parasitic, leaves the mass scale of the dark matter, and hence the energy density ratio Eq.~\eqref{eq:energy}, unexplained. In weakly coupled hidden sector theories, one can set the ADM mass scale by a loop factor below the weak scale~\cite{Cohen:2010kn}, leaving the coincidence as to why the QCD scale is a loop factor below the weak scale.   

In this paper, we propose a mechanism where the dark matter to baryon mass ratio is determined through Dark Unification of the QCD sector with a dark sector.  Both the proton and dark baryon masses are set by confinement in the respective sector, which, because the two sectors are unified at a high scale, occurs at a nearby mass scale in both sectors.  Thus the dark and visible baryon masses are nearby in mass, explaining why the baryon to dark matter {\it mass} density (as opposed to just the {\it number} density) is the same. 

\begin{figure}[t]
\centering
\includegraphics[width=0.5\linewidth]{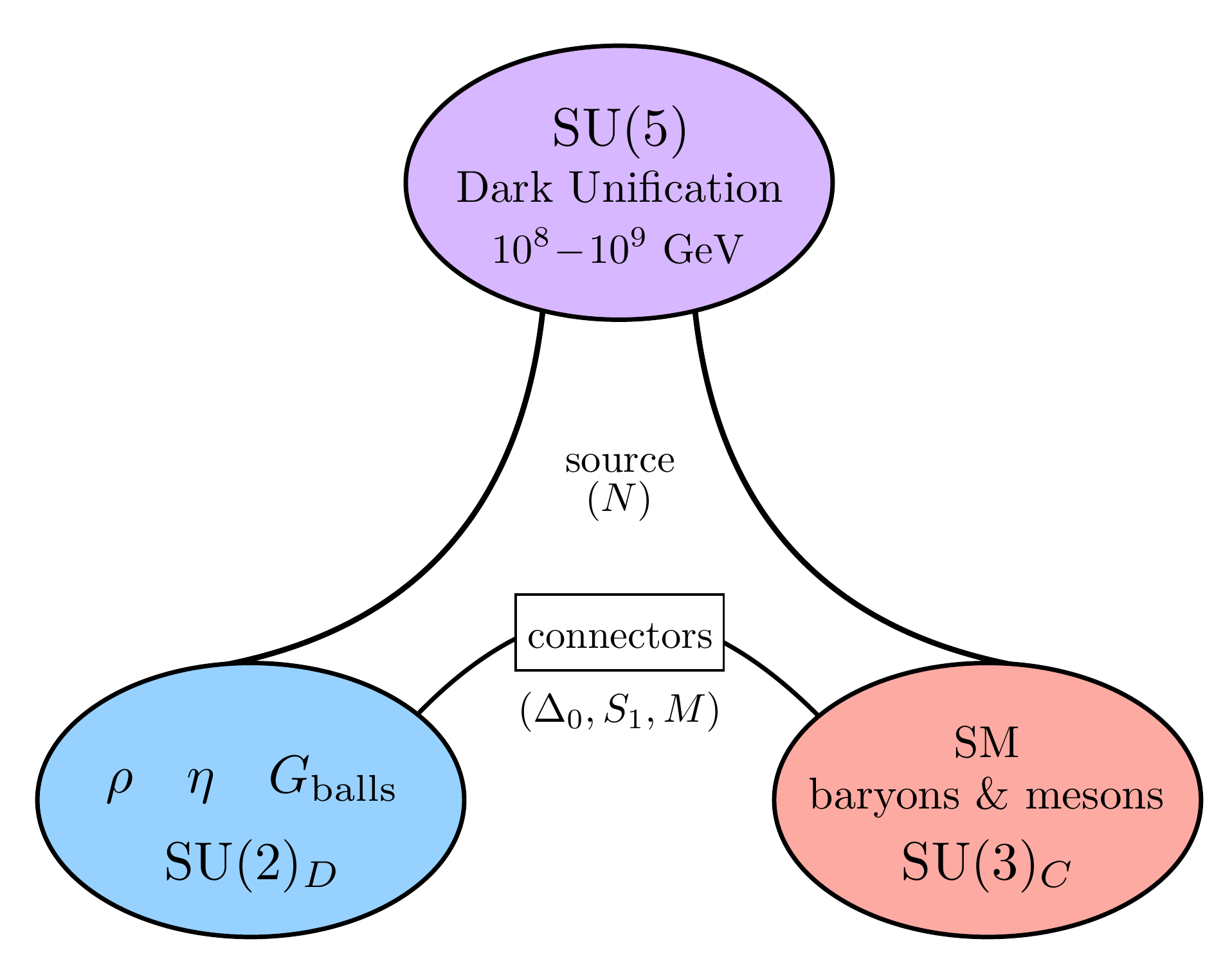} 
\caption{A schematic diagram of Dark Unification, where a dark $\SU(2)_D$ is unified with Standard Model $\SU(3)_C$ into $\SU(5)$ at a scale of $\sim 10^8-10^9$ GeV.  The dark matter in this theory is a dark baryon, $\rho$, part of an isotriplet which is a bound state of two dark quarks (the dark isosinglet is the $\eta$).  The states $M$, $\Delta_0$, the diquark $S_1$ and the sterile neutrinos $N$ appearing from the UV-complete theory play important roles in connecting the dark and visible sectors, mediating dark hadron decays as well as darko-baryo-genesis.}
\label{fig:Messengers}
\end{figure}

We find that the simplest model to realize a Dark-Color Unification Sector (DarCUS) is $\SU(5)$, as shown in Fig.~\ref{fig:Messengers}.  In this theory, $\SU(3)_C \otimes \SU(2)_D$ (plus an additional $\text{U}(1)$) unifies to $\SU(5)$. 
The minimal matter content of the theory is composed of the SM fields, where the quarks live in the fundamental of the extended color group, and a dark sector having a single generation of a vector-like $5$ and $10$.  
While we do not address in detail the further unification of the DarCUS with the Electroweak Theory, we envision that $\SU(5) \otimes \SU(2)_L$ together with the generalized hypercharge could be unified into $\SU(7)$ at a higher scale. We sketch how the main features of the mechanism, as well as the additional matter content required by the generation of the asymmetries, belong to the basic ingredients for the Grand Unified framework near the end of this work.

The unification of the color group with a dark group has been already explored in the literature as a scenario to embed a heavy axion (non-abelian dark group)~\cite{Gherghetta:2016fhp,Gaillard:2018xgk,Gherghetta:2020ofz}, or to embed baryon number (abelian dark group)~\cite{Fornal:2015one,FileviezPerez:2016laj}, while Grand Unified Theories have been generically used to embed dark matter, since the enlarged matter content allows cogeneration of the dark and visible matter asymmetries via global symmetries shared between the sectors~\cite{Barr:2011cz}. However, as far as we are aware, this is the first attempt to connect the proton and dark matter masses directly through the unification of the QCD and a dark confining sector. 

The simple structure of the $\SU(5)$ DarCUS allows for simultaneous generation of the dark matter and baryon asymmetry--darko-baryo-genesis--through the late decay of a Majorana neutrino to both visible and dark states, analogously to the usual leptogenesis~\cite{Fukugita:1986hr}. Although dark baryon number is violated in this process, the decay of the dark matter is suppressed by dimension twelve operators, and therefore consistent with indirect detection constraints. Cosmology predicts a period of Early Matter Domination triggered by the lightest state in the dark spectrum, with reheat temperature between the MeV and GeV scale, and requires that fields carrying color (a colored fermion, a color-adjoint scalar and a scalar diquark) lie near the electroweak scale.

The outline of this paper is as follows. In the next Section, we outline the cast of characters; as in any Grand Unified Theory (GUT), there is a broad range of fields, so we aim to highlight the cast and the roles they play. Then, in Sec.~\ref{sec:theory}, we describe in detail the matter content of the theory and its symmetries. In Sec.~\ref{sec:darkconfinement}, we discuss the resulting low energy spectrum.  Next, we follow the dynamics and cosmological history of the model in Sec.~\ref{sec:cosmology}, focusing on the generation of the dark matter and baryon asymmetries, as well as all relics.  In Sec.~\ref{sec:signatures}, we outline the main observational signatures.  Finally,  in Sec.~\ref{sec:SU7} we discuss how the features of darko-baryo-genesis arise when embedding the DarCUS and $\SU(2)_L$ fields in $\SU(7)$.

\section{Cast of Characters}
\label{sec:COC}
As in any Unified theory, even with few multiplets, there are many fields.  In terms of the relevant {\it dynamics} of the theory, however, only a few of the fields are drivers. Accordingly, our theory features Main Characters, who will be the central focus of our discussion in subsequent sections, as well as Supporting Characters and the Chorus, whose presence is predicted by the minimal $\SU(5)$ theory but are not drivers of the dynamics we study. In Sec.~\ref{sec:SU7} we show how the Main Characters appear automatically when unifying $\SU(5)$ and $\SU(2)_D$ in GUT $\SU(7)$ as the minimal content in the $\bar 7$ and $21$ representations needed to embed the SM fields, while most of the Chorus is not present.
\begin{enumerate}

\item {\bf Main characters}. These fields are so named because they play the dominant roles in the {\em dynamics} of the theory. 
\begin{itemize}
\item The DM candidate ($\chi$): the SM singlet fermion in the fundamental of the dark group that is bound into a dark baryon $\rho$ comprising the dark matter. Its mass will be set by the dark confinement scale.
\item The connectors ($\Delta_0$, $S_1$, $M$): These fields connect the dark and visible baryon sectors and are needed to mediate darko-baryo-genesis. The $\Delta_0$ (scalar) and the visible and dark colored field $M$ (fermion) carry non-zero baryon and dark-baryon number, while $S_1$ is a scalar diquark.
\item The source field ($N$): The field whose late decay generates a baryon and dark baryon asymmetries. A sterile neutrino will play this role analogously to usual leptogenesis.

\end{itemize}
\item {\bf Supporting characters}. These fields mostly appear in the Higgs sectors of the theory and are responsible for the {\em breaking pattern} of $\SU(5)$ to  $\SU(3)_C \otimes \SU(2)_D$.  They will be responsible for the mass spectrum of the theory.
\begin{itemize}
\item $\rho_0$ : A singlet scalar in the $24_H$ breaking $\SU(5) \rightarrow \SU(3)_C \otimes \SU(2)_D \otimes \text{U}(1)_5$. 
\item $\delta_0$: A singlet scalar in the $10_H$ breaking $\text{U}(1)_X \otimes \text{U}(1)_5 \rightarrow \text{U}(1)_Y$.  It is responsible for the splitting between the SM quarks and its partners carrying dark charge, as well as the dark matter and its color partner. 
\item $\rho_8$: A TeV scale color-adjoint, singlet under the dark group, in the $24_H$. It plays an important role in the running of the strong couplings that will determine the ratio between the proton and dark matter masses. 
\end{itemize}

\item {\bf Chorus}.  These fields are important for the {\em mass spectrum} of the theory, though generally do not play a role in the breaking pattern or in the dynamics.
\begin{itemize}
\item New quarks $Q_N$:  while technically residing in the dark sector, these color-carrying quarks split the masses between the dark matter $\chi$ and its colored partner ($q_N$) via the vacuum expectation value (vev) of $\delta_0$.
\item DarCUS scale fields: In this group we have the broken generators (vector darko-quarks $V_\text{DQ}$ and the abelian vector $Z'$), the dark triplet $\rho_3$ living in the $24_H$, and the partners of the SM quarks, which we refer to as {\it quirks}.
\item $\xi$, $q_N$, $\Omega$, $R_D$: these fields, together with $Q_N$, the quirks, and the vector darko-quarks $V_\text{DQ}$, are members of a ${\cal Z}_2$ dark group and carry fractional charges. They can be as heavy as the DarCUS scale. Their abundance will be diluted either by inflation or the annihilation and decay to the SM through their dark hadronization. Note that these fields are absent when embedding $\SU(5)$ and $\SU(2)_L$ in $\SU(7)$.
\end{itemize}
\end{enumerate}

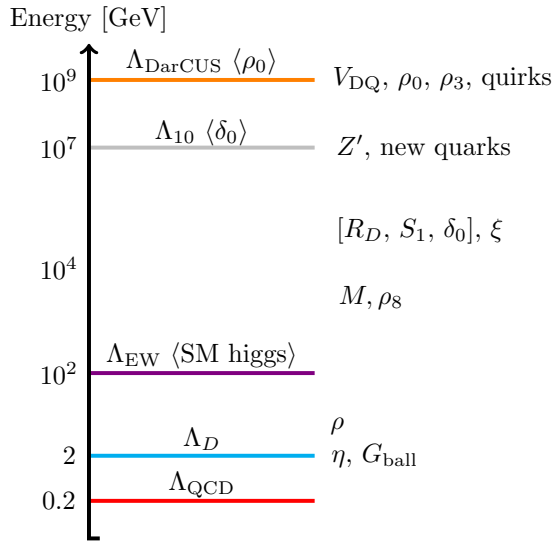
\begin{figure}[t]
\begin{tikzpicture}[line width=1.5 pt,node distance=1 cm and 1.5 cm]

\node (A) at (0, 0) {};
\node (B) at (0, 6.9) {${\rm Energy \ [GeV]}$};
\node (D) at (1.5,2.43) {$\Lambda_\text{EW} \ \langle \text{SM higgs} \rangle$};
\node (E) at (3.3, 1.5) {$\rho$};
\node (Gball) at (3.8, 1.1) {$\eta, \, G_\text{ball}$};
\node (C) at (1.5,6.35) {$\Lambda_\text{DarCUS} \ \langle \rho_0 \rangle$};
\node (F) at (1.5, 5.4) {$\Lambda_{10} \  \langle \delta_0 \rangle$};
\node (G) at (1.5, 1.32) {$\Lambda_D$};
\node (H) at (1.5, 0.72) {$\Lambda_\text{QCD}$};
\node (I) at (3.75, 3.2) {$M, \rho_8$};
\node (Ib) at (4.4, 4.1) {$[R_D, \, S_1, \, \delta_0], \, \xi$};
\node (Hleft) at (-0.4,0.5) {$0.2$};
\node (Gleft) at (-0.26,1.1) {$2$};
\node (Dleft) at (-0.4, 2.2) {$10^2$};
\node (Fleft) at (-0.4,5.2) {$10^7$};
\node (Cleft) at (-0.4,6.1) {$10^9$};
\node (Cright) at (4.7,6.1) {$V_\text{DQ}, \, \rho_0, \, \rho_3, \, \text{quirks}$};
\node (Cright) at (4.45,5.2) {$Z', \,\text{new quarks} $};
\node (AAleft) at (-0.4,3.6) {$10^4$};
\draw[red] (0,0.5) -- (3,0.5); 
\draw[cyan] (0,1.1) -- (3,1.1); 
\draw[violet] (0,2.2) -- (3,2.2); 
\draw[lightgray] (0,5.2) -- (3,5.2); 
\draw[orange] (0,6.1) -- (3,6.1); 
\draw[->, to path={-| (\tikztotarget)}]
  (A) edge (B);
\end{tikzpicture}
\caption{Hierarchies among the field content composing the theory. \label{fig:spectrum}}
\end{figure}

\section{The Theory}
\label{sec:theory}
We will now work through the details of the theory stepping from the high scale of Dark Unification and its breaking, to the matter sector containing the SM, and finally the dark matter sector.  
We summarize the breaking pattern and mass scales of the theory in Fig.~\ref{fig:spectrum}.  We emphasize again that once the multiplets are broken, there are many fields, though the drivers of the dynamics we discuss below will be few; accordingly we emphasize the broad structure of the theory so as not to lose track of the roles the fields play.

\subsection{Dark Unification}

Our starting point is a minimal extension of the SM that contains a non-Abelian dark group:
\begin{equation}
   {\cal G}_5 =  \SU(5) \otimes \SU(2)_L \otimes \text{U}(1)_X.
    \label{eq:GG}
\end{equation}
The breaking pattern of the DarCUS follows\footnote{Such breaking will produce monopoles with mass $\sim \Lambda_\text{DarCUS} / \alpha_\text{DarCUS}$. Kibble's limit~\cite{Kibble:1976sj} estimates a lower bound on the monopole number density per entropy density today $\gtrsim (\Lambda_\text{DarCUS}/M_\text{Pl})^3$. For $\Lambda_\text{DarCUS} \sim 10^{9}$ GeV, it is consistent with astrophysical~\cite{Parker:1970xv} and direct detection constraints~\cite{MACRO:2002jdv}. These monopoles will not survive if inflation occurs at a scale lower than $\Lambda_\text{DarCUS}$.}
\begin{equation}
\begin{split}
&{\cal G}_5 \stackrel{\langle 24_H \rangle}{\to} \SU(3)_C \otimes \SU(2)_L \otimes \text{U}(1)_5 \otimes \text{U}(1)_X \otimes \SU(2)_D \stackrel{\langle 10_H \rangle}{\to} {\cal G}_\text{SM} \otimes \SU(2)_D,
\end{split}
\end{equation}
where ${\cal G}_\text{SM}$ encodes the SM gauge group.\footnote{Below we label the quantum numbers of the fields as $(\SU(3)_C, \SU(2)_L , \text{U}(1)_Y, \SU(2)_D)$ at the level of $ \, {\cal G}_\text{SM} \otimes \SU(2)_D$, while those labelled under the subscript ${\cal G}_5$, refer to the gauge group in Eq.~\eqref{eq:GG}.}
The Higgs sector $24_H$ breaks $\SU(5)$ via the vev of $\rho_0$
\begin{equation}
24_H \sim (24,1,0)_{{\cal G}_5} = \underbrace{(8,1,0,1)}_{\rho_8} \oplus \underbrace{(1,1,0,3)}_{\rho_3} \oplus \underbrace{(3,1,1/6,2)}_{\rho_{(3,2)}} \oplus \underbrace{(\bar 3,1,-1/6,2)}_{\rho_{(3,2)}^c} \oplus \underbrace{(1,1,0,1)}_{\rho_0},
\label{eq:24H}
\end{equation}
while the vev of the hyperchargeless component of $10_H$, {\it i.e.} $\delta_0$, 
\begin{equation}
\begin{split}
    10_H &\sim (10,1,1/5)_{{\cal G}_5} = \underbrace{(\bar 3,1,1/3,1)}_{S_1} \oplus \underbrace{(3,1,1/6,2)}_{R_D} \oplus \underbrace{(1,1,0,1)}_{\delta_0},
    \end{split}
\end{equation}
breaks $\text{U}(1)_5 \otimes \text{U}(1)_X \rightarrow \text{U}(1)_Y$.
The above breaking pattern defines the hypercharge operator as a function of the $\text{U}(1)_X$ charge, $X$, and the non-singular diagonal generator of $\text{SU}(5)$, $T_{24}$, as follows:
\begin{equation}
    Y =X + \frac{1}{\sqrt{15}} \, T_{24}, \quad \text{ where } \quad     T_{24} = \frac{1}{2\sqrt{15}} \, \text{diag} (2,2,2,-3,-3).
\end{equation}
Besides being responsible for breaking $ {\cal G}_5$ down to the SM, we will see that these scalars mainly play a supporting role in the theory: the scalar $S_1$ assists darko-baryo-genesis, while the vev of $\delta_0$ splits the masses of the SM quarks and their dark partners that would otherwise be ruled out by direct detection constraints.

The gauge sector of $\SU(5)$, with analogous quantum numbers under ${\cal G}_\text{SM} \, \otimes \, \SU(2)_D$ as the fields in Eq.~\eqref{eq:24H}, is composed of the eight gluons, three dark gluons and the broken generators $V_\text{DQ} \sim (3,1,1/6,2)$, which we call vector {\it darko-quarks}. Their mass defines the DarCUS scale.

\subsection{Minimal Matter Sector} 

The minimal matter content of the gauge group ${\cal G}_5$ that embeds the SM matter content is
\begin{eqnarray}
 5_q &\sim& (5,2,1/10)_{{\cal G}_5} = \underbrace{(3,2,1/6,1)}_{\displaystyle q} + \underbrace{(1,2,0,2)}_{\displaystyle \Psi}, \\
 \bar 5_u &\sim & (\bar 5,1,-3/5)_{{\cal G}_5} = \underbrace{(\bar 3,1,-2/3,1)}_{\displaystyle u^c} + \underbrace{(1,1,-1/2,2)}_{\displaystyle \eta_u},\\
 \bar 5_d & \sim & (\bar 5,1,2/5)_{{\cal G}_5} =  \underbrace{(\bar 3,1,1/3,1)}_{\displaystyle d^c} + \underbrace{(1,1,1/2,2)}_{\displaystyle \eta_d}.
\end{eqnarray}
We call $\Psi,~\eta_u,$ and $\eta_d$ {\it quirks}, due to their resemblance to the fermion content in Ref.~\cite{Kribs:2009fy}.  We also have the SM Higgs, $H\sim (1,2,1/2)_{{\cal G}_5}$ to break the electroweak symmetry and give mass to the SM fermions through the vev of its neutral component ($v_\text{EW}$), and the SM leptons, which are singlets under the DarCUS symmetry: $\ell \sim (1,2,-1/2)_{{\cal G}_5}$ and $e^c \sim (1,1,1)_{{\cal G}_5}$. In the lepton sector we will also consider at least two sterile neutrinos, $N \sim (1,1,0)_{{\cal G}_5}$, to generate neutrino masses consistently with experiment.  Below we find that these matter fields appear naturally when the DarCUS matter fields are unified with the electroweak fermions into $\SU(7)$.

The minimal matter Lagrangian, consistent with all the symmetries, is
\begin{equation}
    -{\cal L}_{MM} = Y_u \, 5_q H \bar 5_u + Y_d \, 5_q H^\dagger \bar 5_d + Y_e \, \ell H^\dagger e^c + Y_\eta  \, \bar 5_u \bar 5_d 10_H + Y_\Psi  \, 5_q 5_q 10_H^* + Y_\nu \, \ell H N + M_N N N +  \text{h.c.}
    \label{eq:Yukawa1}
\end{equation}
The vev $v_{10} = \langle \delta_0  \rangle \subset 10_H$ generates mass terms for the combinations $\eta_u \, \eta_d$ and $\Psi_u \Psi_d$ (which are the $\SU(2)_L$ components of $\Psi = (\Psi_u, \Psi_d)^T$) and breaks the accidental $\text{U}(1)_Q$ 
(see Sec.~\ref{subsec:Symmetries} for more details) down to a ${\cal Z}_2$. Their mass matrix in the flavor basis reads
\begin{equation}
-{\cal L}_{MM} \supset  \begin{pmatrix} \Psi_u & \eta_d \end{pmatrix} \begin{pmatrix} Y_\Psi \,  v_{10} & Y_u \, v_\text{EW} \\ Y_d^T \, v_\text{EW} & Y_\eta^T \, v_{10} \end{pmatrix}  \begin{pmatrix} \Psi_d \\ \eta_u \end{pmatrix} + \text{h.c.}
\label{eq:quirks}
\end{equation}
A rotation parametrized by $\theta_Q \sim v_\text{EW} / v_{10} \ll 1$ brings the fields to the physical basis, with masses $\sim v_{10}$ for order one Yukawa couplings. After the $\SU(2)_D$ confinement, quirks will form bound states analogous to heavy quarkonia in QCD.

\subsection{Dark Sector}

The theory admits a fermionic Dirac dark matter (DM) candidate, singlet under the SM gauge group, in the fundamental of $\SU(2)_D$. Under the ${\cal G}_5$ gauge group, it will enter as the fundamental of $\SU(5)$ and will come together with a new quark $q_N$, 
\begin{equation}
\begin{split}
&5_\chi \sim (5,1,1/10)_{{\cal G}_5}= \underbrace{(3,1,1/6,1)}_{q_N} \oplus \underbrace{(1,1,0,2)}_{\chi},
\end{split}
\label{eq:5chi}
\end{equation}
and its vector-like partner $\bar 5_\chi \sim (\bar 5,1,-1/10)_{{\cal G}_5}$.  The DM candidate $\chi + \chi^{c}$ will create stable baryons if there is a residual global symmetry protecting their stability.  Since they share the same representation, the fields $q_N$ and $\chi$ are mass degenerate,\footnote{Strictly speaking, the coupling of the vector-like fermions with the adjoint $24_H$ breaks this degeneracy but the size of the mass splitting is proportional to the amount of fine-tuning in the parameters, similarly to the doublet-triplet splitting problem in Georgi-Glashow $\SU(5)$~\cite{Georgi:1974sy}.} but this can be broken with a single copy of 
\begin{equation}
10 \sim  (10,1,-3/10)_{{\cal G}_5} = \underbrace{(\bar 3,1,-1/6,1)}_{Q_N^c} \oplus \underbrace{(3,1,-1/3,2)}_{M} \oplus \underbrace{(1,1,-1/2,1)}_{\xi},
\label{eq:10chi}
\end{equation}
and its vector-like partner $\overline{10} \sim (\overline{10}, 1, 3/10)_{{\cal G}_5}$.

The symmetries allow for interactions involving the matter fields in Eqs.~\eqref{eq:5chi} and~\eqref{eq:10chi} to be 
\begin{equation}
-{\cal L}_{DS} = M_5 \, \bar 5_\chi  5_\chi   + Y_5 \, \bar 5_\chi 24_H 5_\chi  \, + \,  M_{10} \, \overline{10} \, 10  \, +  Y_{10} \, \overline{10} \, 24_H 10 \, +  Y_q \, 5_\chi   10 \, 10_H \epsilon_5  \, +  Y_{\bar q}  \, \bar 5_\chi   \overline{10} \, 10_H^* \epsilon_5   +  \text{h.c.}
\label{eq:Yukawa510}
\end{equation}
In the broken phase the mass matrix for the new quarks reads 
\begin{equation}
{\cal L}_{DS} \supset \begin{pmatrix} Q_N^c &  q_N^c \end{pmatrix} \begin{pmatrix} M_{10} + Y_{10} \, v_{24} /\sqrt{15} & Y_q  \, v_{10} \\ Y_{\bar q} \, v_{10} & M_{5} + Y_5 \, v_{24}/\sqrt{15} \end{pmatrix} \begin{pmatrix} Q_N \\ q_N \end{pmatrix} + \text{h.c.}
\end{equation} 
whereas the dark matter mass is given by $m_\chi = M_5  -3 Y_5 \, v_{24} / (2\sqrt{15})$. We will refer to the mass eigenstates of the above equation as {\it new quarks}.
When $Y_{q} \, v_{10}, Y_{\bar q} \, v_{10}  \gg \Lambda_D \gtrsim m_\chi$, with $\Lambda_D$ the dark color confinement scale, the new quarks can be heavy (at most having a mass on the order of $v_{10}$) consistent with experiment, while the dark matter will be much lighter, having a mass determined by the scale of confinement. From now on we will work in this limit.

\subsection{Accidental symmetries}
\label{subsec:Symmetries}
The gauge symmetries ${\cal G}_5$, together with Lorentz invariance and the matter content, predict three accidental global symmetries, two of which survive when the extended color group spontaneously breaks down to the SM whereas one of them is broken in one unit by the vev of $\delta_0$. The three symmetries can be identified as: 
\begin{itemize}
\item {\bf Baryon number} $\boldmath \textbf{U}(1)_{B}$: The $B$ charges are assigned such that the SM quarks carry 1/3 of baryon charge. 
\item {\bf Dark baryon number} $\boldmath \textbf{U}(1)_{D}$: The $D$ charges are assigned such that the dark matter $\chi$ carries 1/2 of dark baryon charge.
\item {\bf Quirk number} $\boldmath \textbf{U}(1)_Q$: The $Q$ charges are assigned such that a quirk carries 1/2 of quirk number. This symmetry is broken in one unit by the vev of $\delta_0$, becoming an accidental ${\cal Z}_2$ that acts on the fields that carried $\text{U}(1)_Q$ charge (except $\delta_0$).
\end{itemize}
The field interactions dictated by the Lagrangian of the theory fix the baryon, dark baryon and quirk charges of the matter content, which are listed in Tab.~\ref{tab:symmetries}. 
To be more explicit, we expand the interactions coming from the Yukawa couplings $Y_q$, $Y_{\bar q}$, $Y_\eta$ and $Y_\Psi$,
\begin{equation}
\begin{split}
-{\cal L} & \supset Y_\Psi  \left( \Psi  \, \Psi \, \delta_0^* + Q\, Q \, S_1^* + \psi \, Q \,  R_D^* \right)+ Y_\eta  \left( \eta_u \, \eta_d \, \delta_0 + u^c d^c S_1 + \eta_u \, d^c  R_D + \eta_d \, u^c  R_D\right) \\
& \quad + Y_q  \left( q_N \, Q_N^c \, \delta_0 + q_N \, M  \,  R_D + q_N \, \xi \, S_1  + \chi \, Q_N^c \,  R_D + \chi \, M \, S_1 \right) \\
& \quad +  Y_{\bar q}  \left( q_N^c \, Q_N \, \delta_0^* + q_N^c \,  M ^c  \,  R_D^* + q_N^c \, \xi^c \, S_1^*  +  \chi^c \,  Q_N \,  R_D^* +  \chi^c \,  M^c \, S_1^* \right) + \text{h.c.}
\end{split}
\label{eq:YukQuirk}
\end{equation}

\begin{table}
\resizebox{1\textwidth}{!}{
\begin{tabular}{ | c | c | c | c | c | c | c | c | c |  c | c | c | c | c | c | c | }
\hline
 & \multicolumn{2}{|c }{$5_q$} & \multicolumn{2}{|c}{$\bar 5_u$} & \multicolumn{2}{|c}{$\bar 5_d$} & \multicolumn{2}{|c}{$5_\chi$} & \multicolumn{3}{|c}{$10$} & \multicolumn{3}{|c|}{$10_H$} &  $24^\mu$ \\
\hline
\hline
 Sym & $\Psi$ & $q$ & $u^c$ & $\eta_u $ & $d^c $ & $\eta_d $ & $\chi $ & $q_N$ & $Q_N^c$ & $M$ & $\xi$ & $S_1$ & $ R_D$ & $\delta_0$  & $V_\text{DQ}^\mu$  \\
\hline
$U(1)_{B}$ & - & $1/3$ & $-1/3$ & - & $-1/3$ & - & - & $1/3$ & $-1/3$ & $-2/3$ & $-1$  & $2/3$ & $1/3$ & - &  $1/3$  \\
$U(1)_{D}$ & - & - & - & - & - & - & $1/2$ & $1/2$ & $-1/2$ &  $-1/2$ & $-1/2$ & - & - & - & - \\
$U(1)_{Q}$ & $1/2$ & - & - & $-1/2$ & - & $-1/2$ &  - & $-1/2$ & $-1/2$ & - & $1/2$ & -& $1/2$ & $1$ & $-1/2$ \\
\hline
\end{tabular}}
\caption{Accidental symmetries of the theory, $\text{U}(1)_{B}$, $\text{U}(1)_{D}$ and $\text{U}(1)_{Q}$. Once $\delta_0$ gets a vev, the $\text{U}(1)_Q$ is broken to a ${\cal Z}_2$ symmetry acting on the rest of the fields that were charged under $\text{U}(1)_Q$.}
\label{tab:symmetries}
\end{table}

To generate an asymmetry in the color and dark sectors, however, an interaction that breaks the baryon and dark-baryon symmetries is required. We will introduce a scalar in the fundamental of $\SU(5)$,
\begin{equation}
    5_H \sim (5,1,1/10)_{{\cal G}_5} = \underbrace{(1,1,0,2)}_{\Delta_0} \oplus \underbrace{(3,1,1/6,1)}_{\Omega},
    \label{eq:5H}
\end{equation}
to connect the DM and the SM quarks with the sterile neutrinos, whose late decay will source the baryon and dark-baryon asymmetries, as we discuss in detail in the next section. We note that only the field $\Delta_0$ will play a role in darko-baryo-genesis, and that its presence is predicted in $\SU(7)$ since it shares the multiplet with the SM Higgs boson. Similarly, the relevant interactions for darko-baryo-genesis amongst the ones we show below arise from the Yukawa interaction between the minimal representations needed to embed the SM in $\SU(7)$, see Sec.~\ref{sec:SU7} for more details.

The above representation allows for the following interactions:
\begin{equation}
-{\cal L} \supset    Y_{\Delta_0} \, \bar 5_d \, 10 \, 5_H^* + Y_{\bar \chi}  \, \bar 5_\chi \, 5_H  N + Y_{\chi} \, 5_\chi \, 5_H^* N + \text{h.c.},
   \label{eq:5Hinteractions}
\end{equation}
which in terms of the $\, {\cal G}_\text{SM} \otimes \ \SU(2)_D$ fields read
\begin{equation}
\begin{split}
-{\cal L}  \supset & \ Y_{\Delta_0} \, (d^c \, Q_N^c \, \Omega^* +\eta_d \, M \, \Omega^*+ d^c \, M \,  \Delta_0^*  +  \eta_d \, \xi \, \Delta_0^* )  \\
&+ Y_\chi \left(  \chi \, \Delta_0^* + q_N \, \Omega^* \right) N + Y_{\bar \chi}\left( q_N^c \, \Omega + \chi^c \, \Delta_0  \right)N + \text{h.c.}
\label{eq:YukB}
\end{split}
\end{equation}
The interactions weighted by $Y_{\Delta_0}$ define the visible and dark baryon charges of the scalars from the $5_H$,
\begin{equation}
    {\cal Q}_B (\Delta_0) = -1, \quad {\cal Q}_D (\Delta_0) = -1/2, \quad {\cal Q}_B(\Omega) = -2/3, \quad  {\rm and } \quad {\cal Q}_D (\Omega) = -1/2,
    \label{eq:newcharges}
\end{equation}
and from Eq.~\eqref{eq:YukB} one can explicitly see how the interactions with the sterile neutrinos (proportional to $Y_\chi$ and $Y_{\bar \chi}$) break $\text{U}(1)_B$ and $\text{U}(1)_D$ to the diagonal $\text{U}(1)_{B-D}$. The latter is a quasi-conserved symmetry, up to the coupling $\lambda$ from the following quartic interaction
\begin{equation}
V \supset  \lambda \, 10_H^* 5_H 24_H 5_H + \text{h.c.},
\label{eq:lambda}
\end{equation}
which explicitly breaks $\text{U}(1)_{B-D}$. However, the level of breaking can be controlled by the magnitude of the dimensionless parameter $\lambda$, which we will assume to be small, $\lambda \to 0$, for simplicity.\footnote{We have checked that even if $\lambda \sim {\cal O}(1)$, dark and visible baryon number violation constraints are still satisfied due to the highly suppressed operators contributing to the dark and visible baryon decay.} In the following we discuss the repercussion of breaking $\text{U}(1)_B$ and $\text{U}(1)_D$ on the stability of the lightest baryon and dark baryon. 
The simultaneous presence of the interaction weighted by $Y_{\Delta_0}$ and $Y_\chi$ (or $Y_{\bar \chi}$) leads to $\Delta B = 1$ and $\Delta D = 1$, while $\text{U}(1)_{B-D}$ is conserved. Therefore, the lowest dimensional operator for $\rho$ dark matter to decay into a baryon, which must be proportional to both $Y_{\Delta_0}$ and $Y_\chi$ (or $Y_{\bar \chi}$), is
\begin{equation}
\chi \chi QQQ N,
\end{equation}
where $Q$ represents any of the SM quarks. $N$ was added, aside from Lorentz invariance, because the operator must include the interactions weighted by $Y_\chi$ (or $Y_{\bar \chi}$) for visible and dark baryon number violation. 
Because $M_N \gg m_\rho$, $N$ must decay off-shell, $N \to H_0 \nu \to \bar f f \nu$, so that the leading operator is dimension twelve. In Sec.~\ref{sec:cosmology} we will show that it does not threaten the viability of our dark matter candidate. 

The accidental ${\cal Z}_2$, remnant from the $\text{U}(1)_Q$, stabilizes the lightest element of the group composed of $V_\text{DQ}$, $R_D$, $\xi$, $\Omega$, the quirks and the new quarks, which carry fractional charges and part of them visible and/or dark color. Also in Sec.~\ref{sec:cosmology} we discuss why these remnants need not be cosmologically problematic.

\section{Dark Confinement}
\label{sec:darkconfinement}

The matter fields introduced in the previous section, along with the unification structure, fix the unification scale by the ratio of the dark matter to baryon masses.  In this section we fix the DarCUS scale via the observed dark matter to baryon energy density ratio.

\subsection{Scale}

The running of the QCD and $\SU(2)_D$ couplings (with the label S referring to {\it strong}) at the 1-loop level is given by
\begin{equation}
\alpha_\text{S}^{-1}(\mu)= \alpha_\text{S}^{-1}(\Lambda_\text{S}) -  \frac{1}{2\pi} \left( \sum_{i,m_i < \Lambda_S} b_{i} \ln \left(\frac{\mu}{\Lambda_S}\right) + \sum_{i,m_i>\Lambda_S} b_i \ln \left(\frac{\mu}{M_i} \right)\right) \! ,
\end{equation}
with 
\begin{equation}
 b_i = \frac{1}{3} \sum_R S(R) T_i(R) \prod_{j\neq i} \text{dim}_j(R),
 \end{equation}
and $T_i(R)$ is the Dynkin index of the representation, $S(R)=1,2$ and $-11$ for a complex scalar, chiral (two-component) fermion and gauge boson, respectively, and the last term accounts for the multiplicity of the representation under the other gauge groups~\cite{Nath:2006ut}. The sums are separated into fields having mass above or below the strong confinement scale.  Apart from their quantum numbers, the mass hierarchy among the fields in the theory plays a key role in determining the evolution of the running. As is well-known, the unification constraints only depend on the mass-splitting of the representations. In this theory we have a splitting between the different quarks caused by the flavor hierarchies in the Yukawa couplings with the Higgs, as in the SM. We also have a splitting between the quarks and the quirks, and the new quarks and the dark matter, due to the large vev of $\delta_0$. A third splitting occurs between the colored-adjoint $\rho_8$ (TeV scale) and the rest of the $24_H$ fields (DarCUS scale). Such mass-splitting is allowed by the scalar terms in the potential without fine-tuning issues. 

By running the QCD coupling we estimate $\alpha_\text{QCD} (\Lambda_\text{QCD}) \sim 0.8$ at the scale of QCD confinement ($\Lambda_\text{QCD} \sim 330 \text{ MeV}$). We assume that $\SU(2)_D$ confines analogously to QCD when $\alpha_D (\Lambda_D ) \sim \alpha_\text{QCD} (\Lambda_\text{QCD}) \sim 0.8$, which allows us to estimate the dark confinement scale $\Lambda_D$.

\begin{figure}[t]
\includegraphics[width=0.6\linewidth]{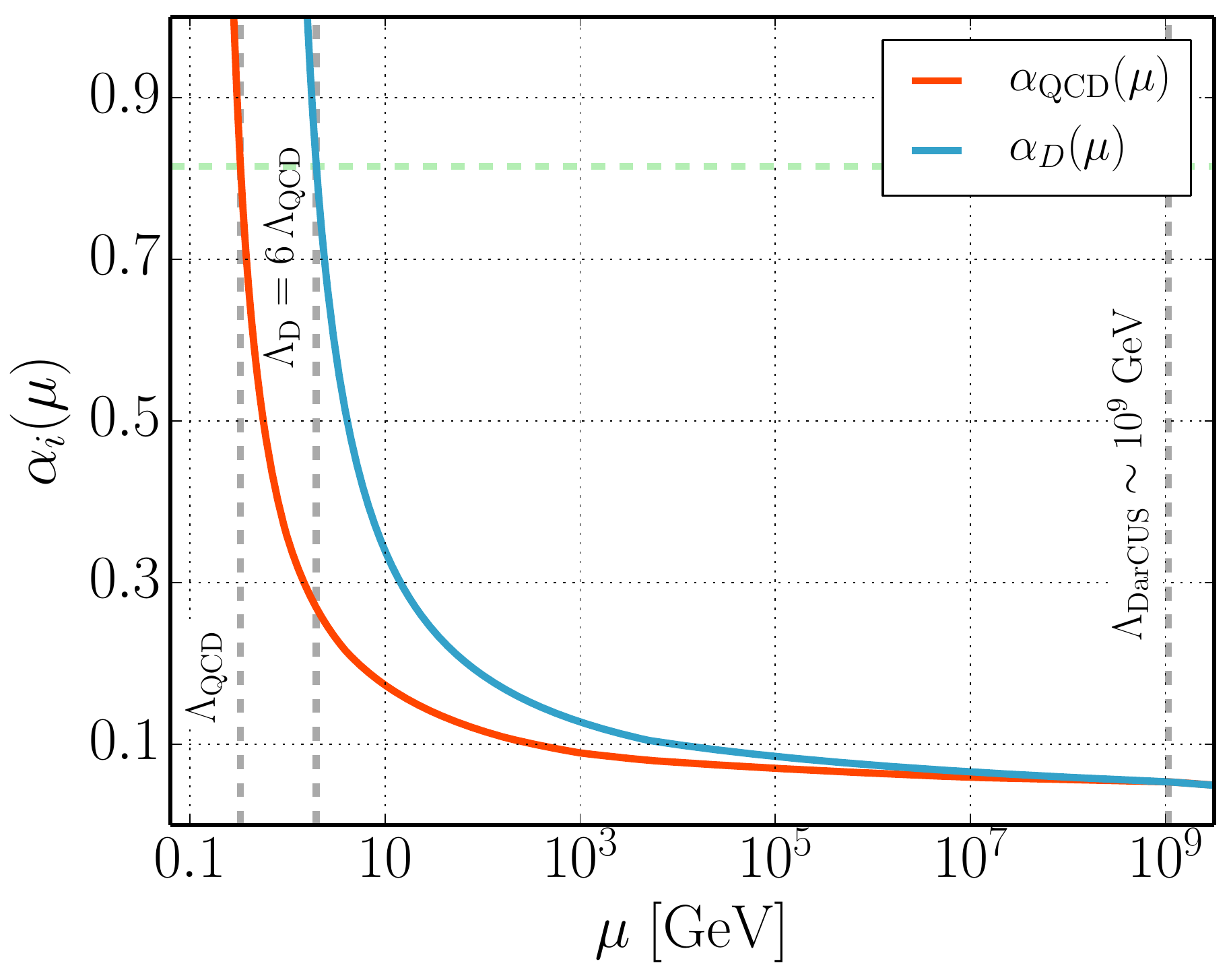} 
\caption{ Running of the strong couplings $\alpha_\text{QCD}(\mu)$ and $\alpha_D(\mu)$, assuming the SM field masses, $m_\chi < \Lambda_D$, $M_{\rho_8} = 1$ TeV, $M_M = 5$ TeV,  $M_{\text{new quarks}} = 10^7$ GeV, and the rest of the fields at $\Lambda_\text{DarCUS} \sim 1 \times 10^9 \text{ GeV}$. The following relation between the confinement scales $\Lambda_\text{D} \sim 6 \, \Lambda_\text{QCD}$ holds.}
\label{fig:running}
\end{figure}
The fields that (a) only contribute to the $\alpha_\text{QCD}$ running are:\footnote{In parenthesis we specify the mass scale only for the fields that belong to a mass-split representation.} gluons, quarks, $\rho_8$ (1 TeV), the scalar diquark $S_1$, the new quarks ($10^7$ GeV), and $\Omega$, (b) only contribute to the $\alpha_\text{D}$ running are: dark gluons, dark matter $\chi$ ($m_\chi < \Lambda_D$), $\rho_3$ ($\Lambda_\text{DarCUS}$), quirks ($\Lambda_\text{DarCUS}$), and $\Delta_0$, (c) contribute to both are: $M$ ($5 \text{ TeV}$), $R_D$ and the vector darkoquarks ($\Lambda_\text{DarCUS}$).  Because $\text{U}(1)_{D-B}$ is approximately a good symmetry, darko-baryo-genesis will require $\Lambda_D \sim 6 \, \Lambda_\text{QCD}$ to reproduce the ratio between the observed baryon and dark matter densities, as it will be discussed in Sec.~\ref{subsec:darkobaryogenesis}. We remark the relevance of the role played by the TeV scale $\rho_8$ to achieve such relation; only looking at the dark and visible colored fields below the confinement scale, the slope for the beta function is $-9/(2\pi) \sim -1.4$ for QCD and $-10/(3\pi) \sim -1.1$ for $\SU(2)_D$. Although they are similar, and therefore their respective couplings are expected to grow close to each other, other {\it light} color degrees of freedom are needed for $\Lambda_D > \Lambda_\text{QCD}$ consistently with dark-color unification. Such requirement is fulfilled in the ${\cal G}_5$ theory since it allows for a TeV scale color octet $\rho_8$.

Using the benchmark mass values indicated in the caption of Fig.~\ref{fig:running} (see Fig.~\ref{fig:spectrum} for a pictorial reference of the mass spectrum adopted) and the aforementioned relation, we find that\footnote{The DarCUS scale could be modified up the amount of lepton asymmetry reprocessed by sphalerons, although in most of the cases will be suppressed due to the low reheat temperatures and the sub-leading branching fraction in the decay channel leading to the lepton asymmetry.}
\begin{equation}
\Lambda_\text{DarCUS} \sim 1 \times 10^9 \text{ GeV}.
\label{eq:scales}
\end{equation}
We stress that the above result depends on the mass spectrum of the theory, and that a lighter mass of the new quarks or a smaller gap between the two confinement scales would lead to a somewhat lower DarCUS scale.

\subsection{Low Energy Spectrum}
\label{subsec:darkbaryonspectrum}

In this theory, $\chi$ is charged under $\SU(2)_D$ and is a singlet under the SM gauge group, and its mass could be below the confinement scale in consistency with the experiment. We focus on this case, $m_\chi < \Lambda_D$. After confinement, $\chi$ will form dark $\SU(2)_D$ composite states. The lightest $SU(2)_D$ singlet state is composed of two fermions $\chi$, which form a spin-1 dark baryon iso-triplet
\beq
\rho = \left( \begin{array}{c} \rho^+ \\ \rho^0 \\ \rho^- \end{array} \right) \sim \left( \begin{array}{c} \chi_D \chi_D \\ \chi_D^c \chi_D \\ \chi^c_D \chi^c_D \end{array} \right).
\eeq
where $\chi_D$ is the four-component Dirac field 
\begin{equation}
\chi_D = \chi_L + \chi_R,
\label{eq:Diracchi}
\end{equation}
where its left-handed projection corresponds to the two-component $\chi$, and the conjugate of its right-projection corresponds to the two-component $\chi^c$,\footnote{In Eq.~\eqref{eq:Diracchi} we have explicitly written the chirality, which was assumed to be left for all two-component fields.} and the dots represent non-renormalizable interactions.
Below we will explain precisely what we mean by the fields to the right of the $\sim$ sign. The $\rho^+,~\rho^-$ both carry dark baryon number,~\footnote{The charge $\pm,0$ here refers to dark baryon charge.} and therefore are a good dark matter candidate. 
The condensate $\langle \chi^c \chi \rangle $ breaks $\SU(2)$ down to $\text{Sp}(2)$ so that there are no broken generators, and hence no dark pions in the theory.  We now sketch in more detail the hadronic spectrum, but see Refs.~\cite{Francis:2018xjd,Kribs:2009fy} for a detailed study.

In four-components, the relevant Lagrangian for $\SU(2)_D$ at low energies is given by
\begin{equation}
{\cal L}_{\SU(2)_D}^\text{eff} = - \frac{1}{2} \text{Tr} \{G_{D\mu \nu} G_D^{\mu \nu} \} + \overline{\chi_D} ( i \slashed{D} - m_\chi) \chi_D + \cdots.
\label{eq:LagChieff}
\end{equation}
By grouping the fields in a doublet of dark baryonic isospin $\SU(2)_{DB}$, 
\begin{equation}
X = \begin{pmatrix} \chi_L \\ -i \sigma_2 C \bar \chi_R^T \end{pmatrix} ,
\end{equation}
we can rewrite the low energy Lagrangian in Eq.~\eqref{eq:LagChieff} as 
\begin{equation}
{\cal L}_{\SU(2)_D}^\text{eff} = \bar X i \slashed{D} X - \frac{m_\chi}{2} \left( X^T i \sigma_2 C E X + \bar X i \sigma_2 C E \bar X^T \right) + \cdots,
\end{equation}
where $E$ is the two-index antisymmetric tensor acting on $\SU(2)_{DB}$, $i \sigma_2$ corresponds to the two-index antisymmetric tensor acting on $\SU(2)_D$ and the charge conjugate operator $C$ acts on the Lorentz space. In the above equation the invariance under $\SU(2)_{DB}$ is explicit. It can also be seen that in the massless case $m_\chi \to 0$, the Lagrangian is invariant under a global phase redefinition, {\it i.e.} $X \to e^{i \theta} X$. Therefore it displays explicitly the enlarged global symmetry coming from the special structure of $\SU(2)$, $\text{U}(2) = \SU(2)_{DB} \times \text{U}(1)_A$ where $\text{U}(1)_A$ is broken by the presence of the mass term (and the EBJ anomaly even if the $\SU(2)_D$ theory has all fermions massless).

Therefore, the lightest state carrying baryon number, which will be part of the following isotriplet ($J^P = 1^-$) ,
\begin{equation}
\rho_\mu^a = \bar X \, \gamma_\mu \tau^a X, 
\end{equation}
where $\tau^a$ are the $a=1\cdots 3$ Pauli matrices acting on isospin indices. Explicitly, 
\begin{eqnarray}
                \rho_\mu^0  & = & \bar X \gamma_\mu \tau^3 X  \qquad  \ \ = \overline{\chi_D} \, \gamma_\mu \chi_D, \\
               \sqrt{2} \rho_\mu^+ & = & \bar X \gamma_\mu (\tau^1 - i \tau^2) X  = \overline{\chi_D^c} \, \gamma_\mu \, i \sigma_2 \, \chi_D, \label{eq:rhoplus}\\
               \sqrt{2}  \rho_\mu^- &= &\bar X \gamma_\mu (\tau^1 + i \tau^2) X  = \overline{\chi_D}  \, \gamma_\mu \, i \sigma_2^T \, \chi_D^c . \label{eq:rhominus}
\end{eqnarray}
The $\rho^+$ baryon (and its antiparticle $\rho^-$) are the dark matter candidates of the theory. 
According to the lattice study in Refs.~\cite{Francis:2016bzf,Francis:2018xjd}, the mass for the lightest baryon $\rho$, $m_\rho \gtrsim 2 \, \Lambda_D$ for any value of the dark quark mass. One of their key results is that the meson $\eta \sim \overline{\chi_D} \, \gamma_5 \, \chi_D$ ($J^P = 0^-$), i.e. the would-be Goldstone boson from $\text{U}(1)_A$, is lighter than the dark matter $\rho$ for any value of $m_\chi$, which allows the DM to annihilate through the channel $\rho \rho \to \eta \eta$. In the following we will adopt $m_\rho \sim  2.5  \, \Lambda_D$ and $m_\eta \sim 2 \, \Lambda_D$ based on their results. 

The light hadronic spectrum also features glueballs, $G_\text{ball}$, isospin and spin singlets with even parity $J^P = 0^+$, {\it i.e.} scalars, which we expect to have mass around $m_{G_\text{ball}} \sim n \, \Lambda_D$, being $n \sim {\cal O}(1)$ parameter.

\section{Cosmology}
\label{sec:cosmology}
We now discuss the components for a successful cosmological evolution, starting with darko-baryo-genesis.  Then we demonstrate that all states, including the dark baryon and glueballs, as well as any charged relics, have an abundance consistent with observation.  This involves ensuring that {\it (i)} darko-baryogenesis occurs naturally with the matter content of the DarCUS; {\it (ii)} the symmetric abundance of dark baryons annihilates sufficiently such that only the asymmetric component sets its abundance; {\it (iii)} the lightest dark hadrons decay early in the Universe; and {\it (iv)} any stable charged relics are consistent with bounds.  We will see that the combination of these requirements places constraints on the mass spectrum of the theory which will in turn generate novel observational signatures, as discussed in the next section.

\subsection{Darko-baryo-genesis}
\label{subsec:darkobaryogenesis}

\begin{figure}[t]
\centering
\begin{equation*}
\begin{gathered}
\resizebox{.25\textwidth}{!}{%
\begin{tikzpicture}[line width=1.5 pt,node distance=1 cm and 1.5 cm]
\coordinate[label = left: $N$] (p1);
\coordinate[right = of p1](p2);
\coordinate[above right = of p2, label=right: $ \Delta_0$] (v1);
\coordinate[below right = of p2,label=right: $ \chi$] (v2);
\draw[fermionnoarrow] (p1) -- (p2);
\draw[scalar] (v1) -- (p2);
\draw[fermion] (p2) -- (v2);
\draw[fill=black] (p2) circle (.07cm);
\end{tikzpicture}}
\end{gathered}
  +   
\begin{gathered}
\resizebox{.32\textwidth}{!}{%
\begin{tikzpicture}[line width=1.5 pt,node distance=1 cm and 1.5 cm]
\coordinate[label = left: $N$] (p1);
\coordinate[right = 0.8cm  of p1](p1a);
\coordinate[right = of p1a](p1b);
\coordinate[right = 0.9 cm of p1b](p2);
\coordinate[above right = of p2, label=right: $ \Delta_0$] (v1);
\coordinate[below right = of p2,label=right: $ \chi$] (v2);
\coordinate[above right = 1cm of p1a,label=$\ell\text{,} \, \chi$](vaux1);
\coordinate[below right = 1cm of p1a,label=$H\text{,} \, \Delta_0$](vaux2);
\coordinate[right=0.5cm of p1b,label=above:$N_{j\neq1}$](vaux3);
\draw[fermionnoarrow] (p1) -- (p1a);
\draw[scalar] (v1) -- (p2);
\draw[fermion] (p2) -- (v2);
\draw[fermionnoarrow] (p1b)--(p2);
\semiloop[fermionnoarrow]{p1a}{p1b}{0};
\semilooptwo[scalarnoarrow]{p1a}{p1b}{0};
\draw[fill=black] (p2) circle (.07cm);
\draw[fill=black] (p1a) circle (.07cm);
\draw[fill=black] (p1b) circle (.07cm);
\end{tikzpicture}}
\end{gathered} %
 +  
\begin{gathered}
\resizebox{.3\textwidth}{!}{%
\begin{tikzpicture}[line width=1.5 pt,node distance=1 cm and 1.5 cm]
\coordinate[label = left: $N$] (p1);
\coordinate[right = of p1](p2);
\coordinate[right=0.55 cm of p2](vmare);
\coordinate[above=0.5cm of vmare,label=$\chi$](vaux1);
\coordinate[below=1.1cm of vmare,label=$\Delta_0$](vaux2);
\coordinate[right= 1cm of vmare,label=right:$N_{j\neq1}$](vaux3);
\coordinate[above right = of p2](v1a);
\coordinate[right = of v1a, label=right: $\Delta_0$] (v1);
\coordinate[below right = of p2](v2a);
\coordinate[right = of v2a,label=right: $ \chi$] (v2);
\draw[fermionnoarrow] (p1) -- (p2);
\draw[scalar] (v1) -- (v1a);
\draw[fermionnoarrow] (v1a)--(v2a);
\draw[fermion](v1a)--(p2);
\draw[scalar] (p2)--(v2a);
\draw[fermion] (v2a) -- (v2);
\draw[fill=black] (p2) circle (.07cm);
\draw[fill=black] (v1a) circle (.07cm);
\draw[fill=black] (v2a) circle (.07cm);
\end{tikzpicture}}
\end{gathered}
\end{equation*}
\caption{Tree and loop-level diagrams contributing to darko-baryo-genesis. }
\label{fig:darkobaryogenesis}
\end{figure}
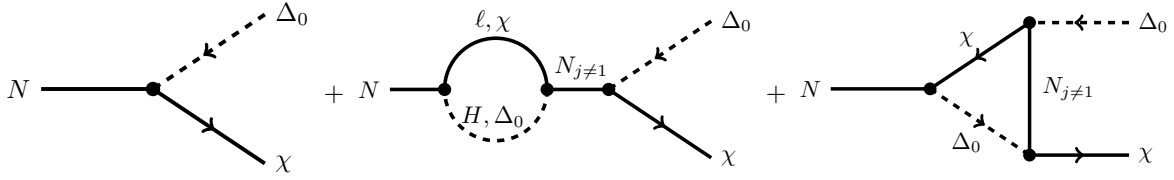

To generate matter and dark matter asymmetries, the three Sakharov conditions~\cite{Sakharov:1967dj} must be fulfilled: the process {\it (i)} must violate baryon $\text{U}(1)_B$ and dark-baryon $\text{U}(1)_D$ numbers, {\it (ii)} must also violate charge conjugation ($C$) and charge conjugation parity ($CP$) symmetries, and {\it (iii)} must occur out of thermal equilibrium. In this theory the sterile neutrinos $N$ are responsible for sourcing the asymmetries, since their late decay can satisfy the aforementioned requirements. We name the process {\it darko-baryo-genesis} because both asymmetries are generated by the same process.\footnote{The source of darko-baryo-genesis in this theory resembles the mechanism proposed in Ref.~\cite{Falkowski:2011xh}. However, in their scenario the generation of the baryon asymmetry is disconnected from the dark matter asymmetry by different decay channels, whereas in our case both asymmetries are co-generated by the same decay branch of $N$.}  The diagrams that generate the needed CP asymmetry are shown in Fig.~\ref{fig:darkobaryogenesis}. 
The late decay of the neutrino preserves $\text{U}(1)_{B-D}$, so that $Y_{\Delta B} = Y_{\Delta D}$ up to the effect of the sphalerons. If the $\Delta_0$ decays to SM quarks before the electroweak phase transition, some of this baryon number will be reprocessed into leptons via the electroweak sphalerons, which we discuss further below. 

The relevant Lagrangian for darko-baryo-genesis in four-component field notation, drawn from the interactions in Eq.~\eqref{eq:YukQuirk},~\eqref{eq:YukB} (in two-components), is given by
\begin{equation}
\begin{split}
    {\cal L} \supset & \quad  \bar u \left(Y_\eta  P_L + Y_\Psi^* P_R \right) d^c S_1 + Y_{\Delta_0} \, \bar d  P_L   M  \Delta_0^* + \bar N \left( Y_\chi^T P_L +  Y_{\bar \chi}^\dagger P_R \right ) \chi \, \Delta_0^* \\
    &+ \overline{\chi^c} \, (Y_q^{\phantom{\dagger}} P_L + Y_{\bar q} \, P_R) \, M \, S_1  + Y_\nu^* \,  \bar \ell \, i \sigma_2  \, H^* P_R  \, N + M_N  \bar N   N +\text{h.c.}
   \end{split}
\end{equation}
Remarkably, the above interactions appear in the Yukawa term arising from the minimal representations needed to embed the SM fields in $\SU(7)$, see Sec.~\ref{sec:SU7}.
In the following we will refer to the above Yukawa couplings as $Y_\text{quirks} \sim {\cal O}(Y_\eta) \sim {\cal O}(Y_\Psi)$, $Y_\text{DN} \sim {\cal O}(Y_\chi) \sim {\cal O}(Y_{\bar \chi})$, and $Y_\text{NewQ} \sim {\cal O}(Y_q) \sim {\cal O}(Y_{\bar q})$ to simplify the discussion.

Given that one key condition for baryo-darko-genesis is the violation of dark baryon number, $\text{U}(1)_D$, we should make sure that $\rho$ is still stable enough to be a good dark matter candidate. Indirect detection bounds on MeV-GeV dark mater constrain the lifetime to $\tau > 10^{24}-10^{28}$ s~\cite{Papucci:2009gd,Cirelli:2009dv,Essig:2013goa} (many orders of magnitude larger than the lifetime of the universe). The decay channel that will dominate the lifetime of $\rho$ ($\tau_\rho$) is given by a dimension-twelve operator, 
\begin{equation}
{\cal O}_{B-D} \sim \frac{Y_f Y_\text{quirks} Y_\nu Y_{\Delta_0} Y_\text{DN} Y_\text{NewQ}}{m_H^2 M_M M_{S_1}^2 M_N M_{\Delta_0}^2} (\bar d^c \chi) (\bar \nu  \, \chi) (\bar d^c u) (\bar f f)
\end{equation}
 which, as discussed in Sec.~\ref{subsec:Symmetries}, comes from the off-shell decay of $N \to H_0^{(*)} \nu \to \nu \bar f f$, where $f(\bar f)$ are light SM fermions and $Y_{f} = m_{f} /v_\text{EW}$ the corresponding Yukawa coupling with the SM Higgs boson. Such an operator leads to the decay of $\rho$ with a lifetime,
       \begin{equation}
       \begin{gathered}
       \begin{tikzpicture}[line width=1.5 pt,node distance=1 cm and 1.5 cm]
       \coordinate[label=left:$\chi$](chi1);
       \coordinate[right = 0.75 cm of chi1](v1);
       \coordinate[above right = of v1](v2);
       \coordinate[below right = of v1](v3);
       \coordinate[above right = 1 cm of v2,label=right:$u$](u);
       \coordinate[below right = 1 cm of v2,label=right:$d^c$](d);
       \coordinate[above right = 1 cm of v3,label=right:$d^c$](d2);
       \coordinate[below left = of v3](v4);
       \coordinate[left = 0.75 cm of v4, label=left:$\chi$](chi2);
       \coordinate[right = of v4](v5);
       \coordinate[above right = 1cm of v5,label=right:$\nu$](ell);
       \coordinate[right =  of v5](vH);
       \coordinate[above right = 1 cm of vH,label=right:$f$](f);
       \coordinate[below right = 1 cm of vH, label=right:$ f$](barf);
       \coordinate[right = 1.2 cm of v1](aux1);
       \coordinate[above = 0.1 cm of aux1, label=$S_1$](S1);
       \coordinate[below = 0.6 cm of aux1, label=$M^c$](M);
       \coordinate[left = 3.2 cm of v3,label=left:$\rho$](rho);
       \coordinate[right = 2 cm of v2,label=below:$\bar n$](n);
       \coordinate[right = 0.6 cm of v4,label=below:$N$](aux2);
       \coordinate[above = 0.5 cm of aux2,label=above:$\Delta_0$](Delta0);
       \coordinate[right = 1.75 cm of aux2,label=below:$H_0$](H);
       \draw[fermion] (chi1)--(v1);
       \draw[fermion] (v1)--(v3);
       \draw[scalar](v2)--(v1);
       \draw[fermion](u)--(v2);
       \draw[fermion](v2)--(d);
       \draw[fermion](v3)--(d2);
       \draw[scalar](v4)--(v3);
       \draw[fermion](chi2)--(v4);
       \draw[fermion](v5)--(ell);
       \draw[scalar](v5)--(vH);
       \draw[fermionnoarrow](v4)--(v5);
       \draw[fermion] (barf)--(vH);
       \draw[fermion] (vH)--(f);
       \draw[fill=black] (v1) circle (.05cm);
       \draw[fill=black] (v2) circle (.05cm);
       \draw[fill=black] (v3) circle (.05cm);
       \draw[fill=black] (v4) circle (.05cm);
        \draw[fill=black] (v5) circle (.05cm);
       \draw[fill=black] (vH) circle (.05cm);
       \draw[fill=GrayLight] (-0.7,-1) ellipse (0.15cm and 1.2cm);
       \draw[fill=GrayLight] (3.8,0.75) ellipse (0.15cm and 1.2 cm);
       \end{tikzpicture}
       \end{gathered}
       \Rightarrow  \tau_\rho^{-1}  \sim  \frac{m_\rho^{17}}{512\pi^5} \frac{|Y_{f}|^2 |Y_\nu|^2 |Y_\text{DN}|^2 |Y_\text{NewQ}|^2 |Y_\text{quirks}|^2 |Y_{\Delta_0}|^2}{m_H^4 M_N^2 M_{\Delta_0}^4  M_M^2  M_{S_1}^4},
       \end{equation}
     where we have approximated the hadronic matrix elements as $\sim {\cal O}(\Lambda_D) \sim m_\rho$.
Even if all Yukawa couplings were order one and the new fields had masses around $5$ TeV, $\tau_\rho$ would be consistent with the indirect detection constraints.

We now discuss the three Sakharov criteria determining the dark baryon asymmetry.
\begin{itemize}
%
\item The {\bf $\boldmath{N}$ lifetime,} or equivalently the reheat temperature $\boldmath T_\text{RH}^N$ associated to its decay, given by
\begin{equation}
T_\text{RH}^{N} \sim \left( \frac{45}{16\pi^3g_*}\right)^{1/4}\sqrt{\Gamma_{N} M_\text{Pl}},
\label{eq:TRH}
\end{equation}
where $M_\text{Pl} \sim 1.2 \times 10^{19}$ GeV.
In the following we assume that the leading contribution to the decay rate of $N$ comes from the channel $N \to \chi \Delta_0^*$, so that the $N$ lifetime is given by
\begin{equation}
\Gamma_{N} \simeq \frac{|Y_\text{DN}|^2}{8\pi} M_N \simeq  6 \times 10^{7} \ \text{ s}^{-1} \ \left( \frac{|Y_\text{DN}|}{10^{-9} }\right)^2 \left( \frac{M_N}{1 \text{ TeV}}\right).
\label{eq:lifetimeN1}
\end{equation}
As Fig.~\ref{fig:SI} shows, $Y_\text{DN} \lesssim 10^{-6}-10^{-4}$ is required for the decay to be out of equilibrium and $Y_\text{DN} \gtrsim 10^{-14}-10^{-12}$ for $N$ to decay before Big Bang Nucleosynthesis (BBN).
\begin{figure}[t]
\includegraphics[width=0.52\linewidth]{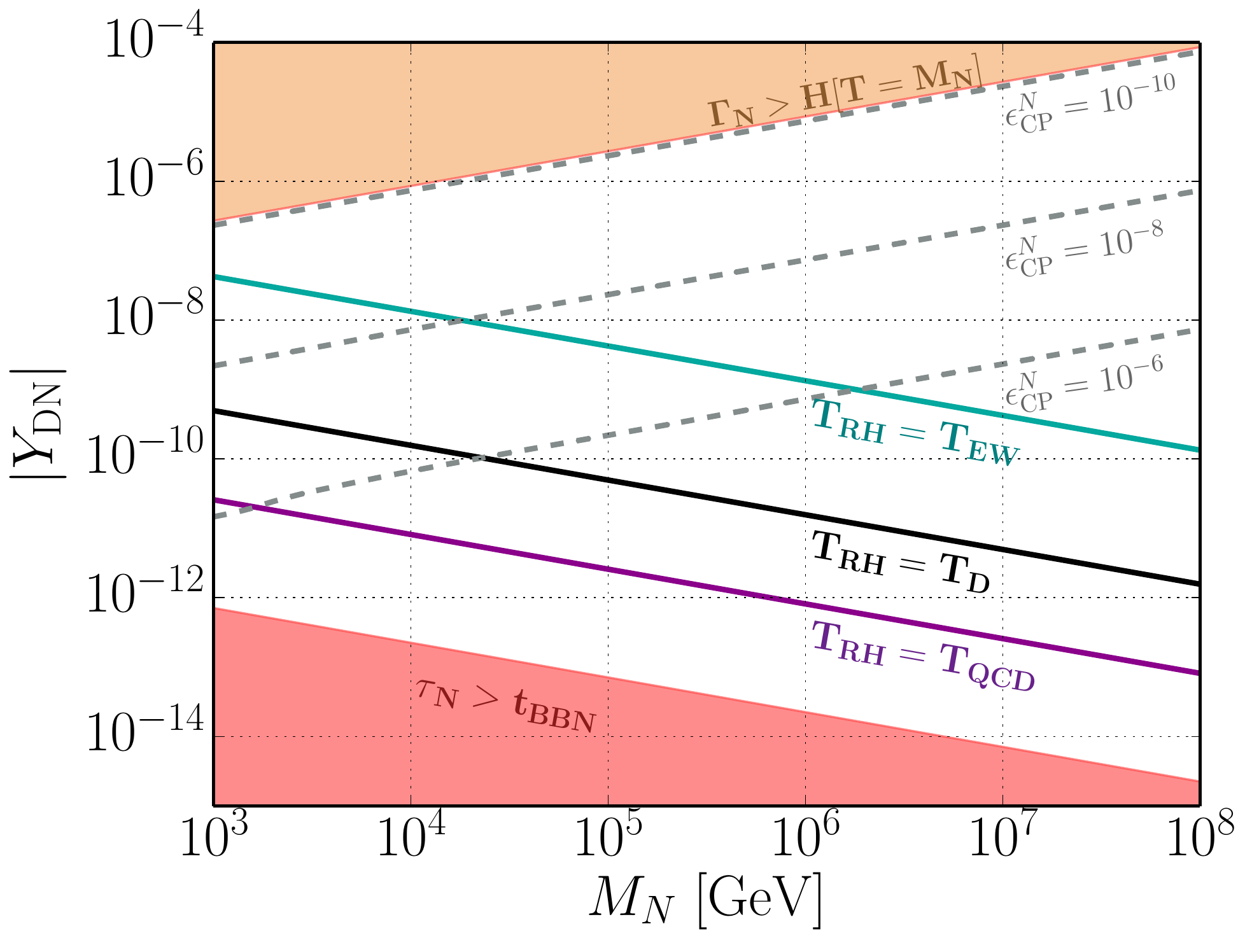} 
\caption{Darko-baryo-genesis through the late decay of $N$ ($M_N > M_{\Delta_0}$). The green, black and purple lines show the parameter space leading to $T_\text{RH}$ equal to the electroweak phase transition ($T_\text{EW} \sim 159 \text{ GeV}$~\cite{DOnofrio:2014rug}), dark phase-transition ($T_D \sim \Lambda_D \sim 2 \text{ GeV}$) and QCD phase transition ($T_\text{QCD} \sim 132 \text{ MeV}$~\cite{HotQCD:2019xnw}), respectively. The red area is ruled out because the Early Matter Domination epoch ends after BBN, while the yellow area is ruled out because the neutrino decays in thermal equilibrium. In grey dashed lines show the amount of CP asymmetry needed to reproduce the observed baryon asymmetry today~\cite{Planck:2018vyg}, assuming that other process that may alter it (reprocess from sphalerons, dilution from early matter domination) are absent.}
\label{fig:SI}
\end{figure}

\item The {\bf $N$ mass}, which is a bare parameter in the Lagrangian and, in principle, totally arbitrary. However, in case inflation takes place at relative low temperatures, we will require that at least the lightest sterile neutrino mass lies below the reheat temperature of inflation (we refer the reader to Sec.~\ref{subsec:ChargedRelics} for more details).

\item A non-zero {\bf CP asymmetry,} $\boldmath \epsilon_\text{CP}^{N}$, in the $N$ decay requires an imaginary contribution from the interference between the tree-level decay process and its 1-loop correction, shown in Fig.~\ref{fig:darkobaryogenesis}. Apart from having complex Yukawa vertices, the optical theorem tells us that in order to obtain an imaginary contribution from the loop function, the propagators must pick up an on-shell contribution within the range of available momenta inside the loop. The presence of more than one sterile neutrino, which is needed to address the observed neutrino oscillations, is crucial to achieve a non-zero $\epsilon_\text{CP}^N$, otherwise the interference term would be proportional to the modulus of the c-number mediating the interaction between $N$ and the SM or the $\SU(2)_D$ fields.

\end{itemize}


The three key elements listed above, $M_{N}$, $T_\text{RH}^N$ and $\epsilon_\text{CP}^N$ will determine the amount of asymmetry generated in the dark and visible baryon sectors through the relation~\cite{Kolb:1988aj}
\begin{equation}
(Y_{\Delta B} =) \, Y_{\Delta D} \simeq \text{Br}(N \to \Delta_0^* \chi) \, \epsilon_\text{CP}^N \frac{T_\text{RH}^N}{M_N},
\label{eq:relAsymmetries}
\end{equation}
neglecting dilution effects due to possible early matter domination and wash-out effects due to scattering, suppressed by two Yukawa couplings squared. The relation between the dark asymmetry and baryon asymmetry, written between brackets, holds up to the effect of the sphalerons, as we discuss below.  
The decay of $N$ generates an equal asymmetry for baryons and dark baryons, which is preserved when there is no violation of baryon number by the electroweak sphalerons. 
Taking $m_p \sim 3 \, \Lambda_\text{QCD}$ and $m_\rho \sim 2.5 \, \Lambda_D$~\cite{Francis:2018xjd}, we find 
 \begin{equation}
 \Lambda_D \sim 6 \, \Lambda_\text{QCD},
 \end{equation} 
 from the observed ratio of dark matter to baryon asymmetries,
$
\Omega_D / \Omega_B = \left( m_\rho \, Y_{\Delta D}\right) \big / \left( m_p \, Y_{\Delta B} \right) \sim 5.
$

The asymmetry generated in $\Delta_0$ can be directly transferred to the baryon sector through the decay 
\begin{equation}
    \begin{gathered}
    \begin{tikzpicture}[line width=1.5 pt,node distance=1 cm and 1.5 cm]
       \coordinate[label= left:$\Delta_0^*$](Delta0);
       \coordinate[right = 1cm of Delta0](v2);
       \coordinate[above right = 1cm of v2,label=right:$d$](dc);
       \coordinate[below right = 1cm of v2](v3);
       \coordinate[above right = 1.25cm of v3](v4);
       \coordinate[below right = 1cm of v3,label=right:$\chi^c$](chi2);
       \coordinate[above right = 1cm of v4, label=right:$d^c$](uc);
       \coordinate[below right = 1cm of v4, label=right:$u$](dc2);
       \coordinate[left = 0.5 cm of v3,label=left:$M$](M);
       \coordinate[right = 0.5 cm of v3,label=right:$S_1$](S1);
       \draw[scalar] (Delta0)--(v2);
       \draw[fermion] (v2)--(dc);
       \draw[fermion] (v3)--(v2);
       \draw[fermion] (chi2)--(v3);
       \draw[scalar] (v3)--(v4);
       \draw[fermion] (uc)--(v4);
       \draw[fermion] (v4)--(dc2);
       \draw[fill=black] (v2) circle (.05cm);
       \draw[fill=black] (v3) circle (.05cm);
       \draw[fill=black] (v4) circle (.05cm);
       \end{tikzpicture}
\end{gathered},
\label{diag:DecayDelta0}
    \end{equation}
if $M_{\Delta_0} < M_M, \, M_{S_1}$. The rate (and lifetime, since the branching fraction $\sim 1$) of the process above is given by
\begin{equation}
\begin{split}
\tau_{\Delta_0}^{-1} \simeq & \frac{|Y_\text{NewQ}|^2|Y_{\Delta_0}|^2|Y_\text{quirks}|^2}{1024 \pi^5} \frac{M_{\Delta_0}^7}{M_S^4 M_M^2} \\
& \simeq 2.5 \times 10^{14} \text{ s}^{-1} \left(\frac{|Y_\text{NewQ}|}{1}\right)^2 \left(\frac{|Y_{\Delta_0}|}{0.01}\right)^2 \!\! \left(\frac{|Y_\text{quirks}|}{1}\right)^2 \!\! \left(\frac{M_{\Delta_0}}{2\text{ TeV}}\right)^7 \!\! \left( \frac{10  \text{ TeV}}{M_{S_1}}\right)^4 \!\! \left(\frac{5 \text{ TeV}}{M_M}\right)^2 \!\! ,
\end{split}
\label{eq:Delta0lifetime}
\end{equation}
which fixes the reheat temperature as given in Eq.~\eqref{eq:TRH} (substituting $\Gamma_{\Delta_0}$ instead). If instead $\Delta_0$ decays to an on-shell $M$ or $S_1$, the asymmetry will be step-by-step transferred to the SM baryons. If the lowest reheat temperature of the decays transferring the asymmetry to the baryon sector is above the electroweak scale, the effect of the sphalerons distributing the asymmetry in the SM fermions reduces the baryon asymmetry to $Y_{\Delta B} \sim 0.36 \, Y_{\Delta D}$~\cite{Harvey:1990qw}, implying a modified DarCUS scale of $\Lambda_\text{DarCUS} \sim 6 \times 10^7$ GeV.

We note that the entropy injected in the plasma from the (potential) early matter domination period induced by the long-lived dark hadrons could dilute both dark and visible baryon asymmetries, depending on the reheat temperature of the latest decay transferring the asymmetry to the baryon sector. In this case, a larger CP asymmetry than that shown in Fig.~\ref{fig:SI} would be needed to reproduce the observed baryon asymmetry. 

\subsection{Annihilation of the symmetric abundances in the dark sector}
%
Below the confinement scale, the dark composite hadrons will interact amongst themselves with geometric cross-sections $\sigma \sim 4\pi / \Lambda_D^2$. From the mass spectrum computed in Ref.~\cite{Francis:2018xjd}, unlike in QCD, the pseudoscalar meson $\eta$ is lighter than the DM candidate $\rho$ for any mass of the dark quark $\chi$. Therefore, the symmetric component of the dark matter ({\it i.e.} stable states $\rho^+$ and $\rho^-$, see Eqs.~\eqref{eq:rhoplus} and~\eqref{eq:rhominus}) will annihilate to a pair of glueballs (if $m_{G_\text{ball}} \lesssim m_\rho$), or to a pair of $\eta$ fields with cross-section (using $m_\rho \sim 2.5\, \Lambda_D$),
\begin{equation}
\langle \sigma_\text{ann} v \rangle \sim  \left(\frac{4\pi}{\Lambda_D^2}\right) \left(\frac{3}{2}\frac{\Lambda_D}{m_\rho}\right)^{1/2} \sim 4 \times 10^{-17} \text{cm}^3 \text{ s}^{-1} \left(\frac{2 \text{ GeV}}{\Lambda_D}\right)^{2},
\end{equation}
which is adequate to remove the symmetric abundance, as found Ref.~\cite{Lin:2011gj}.

Furthermore, we should ensure that the lightest non-stable field in the dark spectrum decays to the SM with a lifetime shorter than one second in order to be consistent with BBN constraints; this state could be either the $\eta$ or the glueballs, so we will check that both states can decay sufficiently fast.  The dark gluons ($g_D$) confine to form glueballs ($G_\text{ball}$) below the dark confinement scale, acquiring a mass $m_{G_\text{ball}} \sim n \, \Lambda_D$, where $n > 1$. In QCD, glueballs have masses about $3-5 \, \Lambda_\text{QCD}$~\cite{Ochs:2013gi}, suggesting that the $\eta$ is more likely to be the lightest unstable state. However, since a lattice study of the glueballs in $\SU(2)_D$ with one flavor would be needed to estimate the parameter $n$, we check that its lifetime to decay to a pair of gluons is sufficiently short via the process: 
\begin{equation}
\begin{gathered}
\begin{tikzpicture}[line width=1.5 pt,node distance=1 cm and 1 cm]
\coordinate[label=left:$g_D$] (i1);
\coordinate[right = 0.7 cm  of i1](v1);
\coordinate[below = 1.5 cm of v1](v2);
\coordinate[left = 0.7cm of v2,label=left:$g_D$](i2);
\coordinate[right = 1.75 cm of v1](v3);
\coordinate[right = of v3,label=right:$g$](f3);
\coordinate[below = 1.5 cm of v3](v4);
\coordinate[right = of v4,label=right:$g$](f4);
\coordinate[right = 0.65cm of v1](aux1);
\coordinate[below = 1.05 cm of aux1,label=$\, \, \, \, \,\, \, M\text{, }R_D$](M);
\coordinate[left = 3 cm of M,label=$G_\text{ball}$];
\draw[gluon](i1)--(v1);
\draw[gluon](i2)--(v2);
\draw[gluon](v3)--(f3);
\draw[gluon](v4)--(f4);
\draw[fermion](v1) -- (v3);
\draw[fermion](v3)--(v4);
\draw[fermion](v4)--(v2);
\draw[fermion](v2)--(v1);
\draw[fill=black] (v1) circle (.05cm);
\draw[fill=black] (v2) circle (.05cm);
\draw[fill=black] (v3) circle (.05cm);
\draw[fill=black] (v4) circle (.05cm);
\draw[fill=GrayLight] (-0.9,-0.75) ellipse (0.15cm and 1cm);
\end{tikzpicture}
\end{gathered},
\label{eq:glueballdecay}
\end{equation}
corresponding to a dimension-eight effective operator, 
\begin{equation}
\sim \, \left( \frac{1}{4} \delta_{ab} \delta_{cd} \right) \frac{\displaystyle \alpha_D \, \alpha_\text{QCD}}{\displaystyle M_{M}^4}G_{D\mu \nu}^aG_D^{\mu \nu,b} G_{\alpha \beta}^cG^{\alpha \beta,d},
\end{equation}
where $a,b=1... 3$ and $c,d = 1... 8$. The decay rate is
\begin{equation}
\Gamma_{G_\text{ball} \to g g} \simeq \frac{1}{8\pi} \frac{1}{6}\frac{\alpha_D^2 \alpha_\text{QCD}^2}{M_{M}^8}(n \, \Lambda_D)^9,
\label{eq:GammaGB}
\end{equation}
corresponding to a lifetime, for $n=1$, 
\begin{equation}
\tau_{G_\text{ball}} \simeq 0.5 \text{ s}  \left(\frac{0.8}{\alpha_\text{QCD}}\right)^2 \left(\frac{0.8}{\alpha_D}\right)^2 \left(\frac{M_{M}}{1 \text{ TeV}}\right)^8 \left(\frac{2 \text{ GeV}}{\Lambda_D}\right)^9.
\end{equation}
An equivalent contribution comes from $R_D$. Requiring the glueballs to decay before BBN, in the event that the glueballs are the lightest unstable confined dark state, sets a constraint on the mass of either of $M$ or $R_D$ to be $ \lesssim n \text{ TeV}$.
In the more likely case that $\eta$ is the lightest hadron from the dark spectrum, whose mass is $m_\eta \sim 2 \, \Lambda_D$~\cite{Francis:2018xjd}, its decay is given by
\begin{equation}
\begin{gathered}
\begin{tikzpicture}[line width=1.5 pt,node distance=1 cm and 1 cm]
\coordinate[label=left:$\chi$] (i1);
\coordinate[right = 0.75 cm of i1](v1);
\coordinate[below = 1.5 cm of v1](v2);
\coordinate[left = 0.75 cm of v2,label=left:$ \chi$](i2);
\coordinate[below right = 1.25 cm of v1](v3);
\coordinate[above right = 1.25 cm of v3,label=right:$g$](g1);
\coordinate[below right = 1.25 cm of v3,label=right:$g$](g2);
\coordinate[below = 0.75cm of i1](aux);
\coordinate[left = 0.9 cm of aux,label=left:$\eta$](eta);
\coordinate[above=0.2cm of aux](aux1);
\coordinate[right = 0 cm of aux,label=right:$M$](M);
\coordinate[right = 1.5 cm of aux1,label=above:$\,\,\,S_1$](aux2);
\draw[fermion](i1)--(v1);
\draw[fermion](v1)--(v2);
\draw[fermion](v2)--(i2);
\draw[gluon](v3)--(g1);
\draw[gluon](v3)--(g2);
\draw[scalar](v3) -- (v1);
\draw[scalar](v2)--(v3);
\draw[fill=black] (v1) circle (.05cm);
\draw[fill=black] (v2) circle (.05cm);
\draw[fill=black] (v3) circle (.05cm);
\draw[fill=GrayLight] (-0.65,-0.75) ellipse (0.15cm and 1cm);
\end{tikzpicture}
\end{gathered} +
\quad
\begin{gathered}
\begin{tikzpicture}[line width=1.5 pt,node distance=1 cm and 1 cm]
\coordinate[label=left:$\chi$] (i1);
\coordinate[right = 0.75 cm of i1](v1);
\coordinate[below = 1.5 cm of v1](v2);
\coordinate[left = 0.75 cm of v2,label=left:$ \chi$](i2);
\coordinate[right = 1.5 cm of v1](v3);
\coordinate[right = 1.5 cm of v2](v4);
\coordinate[right = 0.75 cm of v3,label=right:$g$](g1);
\coordinate[right = 0.75 cm of v4,label=right:$g$](g2);
\coordinate[below = 0.75cm of i1](aux);
\coordinate[left = 0.9 cm of aux,label=left:$\eta$](eta);
\coordinate[right = 0 cm of aux,label=right:$M$](eta);
\coordinate[right = 1.5 cm of aux,label=right:$S_1$](eta);
\draw[fermion](i1)--(v1);
\draw[fermion](v1)--(v2);
\draw[fermion](v2)--(i2);
\draw[gluon](v3)--(g1);
\draw[gluon](v4)--(g2);
\draw[scalar](v3) -- (v1);
\draw[scalar](v4)--(v3);
\draw[scalar](v2)--(v4);
\draw[fill=black] (v1) circle (.05cm);
\draw[fill=black] (v2) circle (.05cm);
\draw[fill=black] (v3) circle (.05cm);
\draw[fill=black] (v4) circle (.05cm);
\draw[fill=GrayLight] (-0.65,-0.75) ellipse (0.15cm and 1cm);
\end{tikzpicture}
\end{gathered} +
\quad
\begin{gathered}
\begin{tikzpicture}[line width=1.5 pt,node distance=1 cm and 1 cm]
\coordinate[label=left:$\chi$] (i1);
\coordinate[right = 0.75 cm of i1](v1);
\coordinate[below = 1.5 cm of v1](v2);
\coordinate[left = 0.75 cm of v2,label=left:$ \chi$](i2);
\coordinate[right = 1.5 cm of v1](v3);
\coordinate[right = 1.5 cm of v2](v4);
\coordinate[right = 0.75 cm of v3,label=right:$g$](g1);
\coordinate[right = 0.75 cm of v4,label=right:$g$](g2);
\coordinate[below = 0.75cm of i1](aux);
\coordinate[left = 0.9 cm of aux,label=left:$\eta$](eta);
\coordinate[right = 1.5 cm of aux,label=right:$M$](eta);
\coordinate[right = 0 cm of aux,label=right:$S_1$](eta);
\draw[fermion](i1)--(v1);
\draw[scalar](v2)--(v1);
\draw[fermion](v2)--(i2);
\draw[gluon](v3)--(g1);
\draw[gluon](v4)--(g2);
\draw[fermion](v1) -- (v3);
\draw[fermion](v3)--(v4);
\draw[fermion](v4)--(v2);
\draw[fill=black] (v1) circle (.05cm);
\draw[fill=black] (v2) circle (.05cm);
\draw[fill=black] (v3) circle (.05cm);
\draw[fill=black] (v4) circle (.05cm);
\draw[fill=GrayLight] (-0.65,-0.75) ellipse (0.15cm and 1cm);
\end{tikzpicture}
\end{gathered} \! ,
\label{eq:etadecay}
\end{equation}
corresponding to dimension-seven effective operators, for example
\begin{equation}
\sim \delta_{ij} \delta_{ab}  \frac{\alpha_\text{QCD}}{4 \pi}\frac{Y_\text{NewQ}^2}{M_{S_1/M}^3} \,  \bar \chi_i \gamma_5 \chi_j \, G_{\mu \nu}^a \tilde G^{\mu \nu,b} ,
\end{equation}
where $i,j=1,2$ are $\SU(2)_D$ indices, whereas $a,b=1...8$ are the number of gluons. Thus, the lifetime is given by
\begin{equation}
\tau_\eta^{-1}  \sim \frac{1}{4\pi} \frac{\alpha_\text{QCD}^2}{(4\pi)^2}\frac{|Y_\text{NewQ}|^4}{M_{S_1/M}^6} \, m_\eta^7 = 8 \text{ s}^{-1} \left(\frac{\alpha_\text{QCD}}{0.8}\right)^2 \left(\frac{|Y_\text{NewQ}|}{1}\right)^4 \left(\frac{10 \text{ TeV}}{M_{S_1/M}}\right)^6 \left(\frac{m_\eta}{4 \text{ GeV}}\right)^7.
\label{eq:etalifetime}
\end{equation}
The dominant contribution to the $\eta$ lifetime will come from the heavier of the two connectors, $M$ or $S_1$. Hence, we expect both $M_M$ and $M_{S_1}$ below the $10$ TeV scale. If the baryon asymmetry is reprocessed by the sphalerons, {\it i.e.} the lowest reheat temperature involved in the transfer or generation of the $\Delta_0$ asymmetry to the baryon sector $\geq T_\text{EW}$, then $\Lambda_D \sim 0.7$ GeV, and the masses for $M$ and $S_1$ must be lighter than $\sim 3$ TeV.

To summarize, the rest of dark hadrons remain in thermal equilibrium with the lightest dark hadron at $T \simeq \Lambda_D$, and the lightest dark hadron decays before BBN. Thus, their abundance is removed through the decay of $\eta$ or the glueballs, similar to the models of Asymmetric Dark Matter described in Ref.~\cite{Morrissey:2009ur,Cohen:2010kn}.  Note that the decay of the lightest dark hadron requires $M,~S_1$ or $R_D$ to have a mass in the 1-10 TeV range, implying possible production at colliders.  We discuss this below in Sec.~\ref{sec:signatures}.

\subsection{Charged relics}
\label{subsec:ChargedRelics}
The remnant ${\cal Z}_2$ symmetry from the spontaneous symmetry breaking of $\text{U}(1)_Q$ leaves a fractionally charged relic amongst the states carrying the ${\cal Z}_2$.  Before proceeding further, we should emphasize that the presence of these charged relics depends on how the $\SU(5)$ is UV-completed; for example, if $\SU(5)$ is unified together with $\SU(2)_L$ into $\SU(7)$, the accidental ${\cal Z}_2$ symmetry no longer exists and charged relics are no longer a consideration.  However, to emphasize that the theory can still be consistent with observation, we briefly discuss how the charged relics can be suppressed.

Explicitly, the fields, their quantum numbers, and the multiplet from which they originate, are
\begin{eqnarray*}
&&\underbrace{\eta_d \sim (1,1,1/2,2), \ \eta_u \sim (1,1,-1/2,2), \ \Psi \sim (1,2,0,2)}_{\bar 5_d, \ \bar 5_u, \ 5_q \ (\text{quirks})}, \ \underbrace{q_N \sim (3,1,1/6,1)}_{5_\chi}, \ \underbrace{q_N^c \sim (\bar 3,1,-1/6,1)}_{\bar 5_\chi},\\
&&  \underbrace{Q_N^c \sim (\bar 3,1,-1/6,1), \  \xi \sim (1,1,-1/2,1)}_{10}, \  \underbrace{Q_N \sim (3,1,1/6,1), \  \xi^c \sim (1,1,1/2,1)}_{\overline{10}},  \ \underbrace{R_D \sim (3,1,1/6,2)}_{10_H}\\
&& \underbrace{\Omega \sim (3,1,1/6,1)}_{5_H}, \ \underbrace{V_\text{DQ} \sim (3,1,1/6,2), \ V_\text{DQ}^c \sim (\bar 3,1,-1/6,2)}_{\text{vector darko-quarks}}. \\
\end{eqnarray*}

There are multiple mechanisms to dramatically reduce the relic abundance:
\begin{itemize}

\item {\bf Inflation}. If the stable field carrying a fractional charge is heavier than the inflationary scale, its abundance is erased by inflation. 
In this scenario either {\it (i)} inflation occurs at low scales (below $\sim 10$ TeV), or {\it (ii)} we split the mass of the connectors $M$ and $S_1$ from their ${\cal Z}_2$-charged partner belonging to the same $\SU(5)$ representation.

\item {\bf Annihilation}. 
At temperatures around the dark and color confinement scales, the relic ${\cal Z}_2$-charged fields hadronize with the SM fields and DM to form hybrid stable hadrons. Since the lightest state ${\cal Z}_2$ group is generally in the fundamental of $\SU(3)_C$ or/and $\SU(2)_D$, the hybrids have large geometric cross-sections, which leads to a recoupling effect that further suppresses their abundance via an interaction cross-section consistent with the QCD scale:
\begin{equation}
\Omega_{R_D}^\text{hybrid} h^2 \sim  2 \times 10^{-11} \,  \frac{M_{R_D} \sqrt{g_*}}{\Lambda_\text{QCD}} \, \left(  \frac{\text{GeV}^{-2}}{\langle \sigma v \rangle} \right) \simeq 
3 \times 10^{-7} \, \left(\frac{M_{R_D}}{3 \text{ TeV}}\right)^{3/2}\left( \frac{g_*}{10}\right)^{1/2} .
\label{eq:hybridrelic}
\end{equation}
The above abundance is likely to be further diluted due to early matter domination triggered by the late decay of the neutrino, the connectors, or the lightest dark hadron, as discussed in the previous section. 
This is a very small residual relic abundance of the hybrids, but nevertheless they are potentially subject to strong bounds because of their strong and/or electromagnetic interactions with matter. 
If the hybrids are captured by SM nucleons via the Coulomb or/and strong force, they behave like heavy baryons. An abundance $\Omega_{R_D}^\text{hybrids} h^2 < 0.0044$ is consistent with CMB constraints~\cite{McDermott:2010pa,Langacker:2011db}. Since they are expected to be stopped at the crust of the Earth or in meteorites, experiments which search for fractionally charged states in the rock, based on atomic mass spectroscopy~\cite{Nitz:1986gb}, Rutherford backscattering of heavy ions~\cite{Polikanov:1990sf}, as well as levitometer and Millikan liquid drop methods (see Ref.~\cite{Perl2009} for a review on fractional charged particles), may apply. However, these bounds depend strongly on the capture rate by a nucleon, which is very uncertain, and whether the analyzed material stores a cosmologically representative amount of these hybrids.

To summarize, there are multiple avenues by which the charged relic can be consistent with cosmological and terrestrial bounds.  In addition, the ${\cal Z}_2$ group is absent when the DarCUS is embedded within the SM $\SU(2)_L$ in a GUT (see Sec.~\ref{sec:SU7}), such that charged relics may be absent when the theory is fully UV completed.

\end{itemize}

\section{Signatures}
\label{sec:signatures}

Our model generically features a wide range of astrophysical and phenomenological signatures that we briefly summarize here, leaving detailed study for future work.

\subsection{Phenomenological}

\begin{itemize}  

\item {\bf Collider Signatures.}  The top-down structure of the theory with a hidden sector unified with the Standard Model at a high scale echoes the Hidden Valley.  The connectors $M$ and $S_1$ (or $M$ or $R_D$ if the lightest dark states are the dark glueballs), all of which are colored, can be pair-produced at a collider, see Fig.~\ref{fig:PairProduction}.  Their decay channels can be inferred from Eq.~\eqref{diag:DecayDelta0}. 
Both $M$ and $S_1$ may decay promptly with lifetimes
\begin{equation}
\begin{split}
\tau_M^{-1} \simeq &\frac{1}{128 \pi^3} \left| Y_\text{NewQ}\right |^2 \left | Y_{\text{quirks}} \right |^2 \frac{M_M^5}{M_{S_1}^4}  \\
&\simeq  9.3 \times 10^{21} \text{ s}^{-1} \left| \frac{Y_\text{NewQ}}{1}\right|^2 \left|\frac{Y_{\text{quirks}}}{1}\right|^2 \left(\frac{M_M}{3\text{ TeV}}\right)^5 \left(\frac{10 \text{ TeV}}{M_{S_1}}\right)^4,
\end{split}
\label{eq:Mlifetime}
\end{equation}
and
\begin{equation}
\tau_{S_1}^{-1} \simeq \frac{M_{S_1}}{4\pi} |Y_\text{quirks}|^2 \simeq 1.2 \times 10^{27} \text{ s}^{-1} \left(\frac{M_{S_1}}{10 \text{ TeV}}\right) \left( \frac{|Y_\text{quirks}|}{1}\right)^2.
\end{equation}
$M$ decays to quarks and the dark field $\chi$, which could hadronize into stable dark baryons plus unstable dark hadrons like the $\eta$ that later decay back to the SM. Since the unstable dark hadron lifetimes range between $\sim {\cal O}(10^{-6}) - {\cal O}(1)$ s, we expect them to decay outside the detector, although in some cases one might have displaced vertices. 
The decay $M \to 2 \text{ JET}s + \text{MET}$ resembles some gluino searches, allowing one to translate the existing SUSY bounds to constraints on $M$. Using Run-II data, the ATLAS~\cite{ATLAS:2017tny,ATLAS:2018lob,ATLAS:2018yey,ATLAS:2020syg} and CMS~\cite{CMS:2017kku,CMS:2016kce,CMS:2018qxv} collaborations rule out gluino masses below $2-2.4$ TeV, depending on the gluino lifetime. 
Assuming that the t-channel dominates the pair-production cross-section, then $\sigma (gg \to \tilde g \tilde g) =(27/4) \, \sigma(gg \to M \bar M)$, such that we expect the aforementioned bounds on the gluinos to be slightly weaker for $M$. 
Scalar diquark constraints at the LHC place a strong lower bound on the $S_1$ mass, $M_{S_1} > 7.5$ TeV, for degenerate electromagnetic strength couplings~\cite{CMS:2019gwf}, via off-shell pair-production of the diquark, with each decaying to a pair of jets. The bound could be relaxed assuming a non-trivial flavor structure of the $Y_\eta$ and $Y_\Psi$ Yukawa matrices. On the other hand, the production cross-section for masses of $S_1$ below $\sim 1 $ TeV is dominated by the gluon fusion and quark anti-quark annihilation~\cite{ATLAS:2015hsi}, which is independent on the diquark coupling. Thus, we expect $M_{S_1} \gtrsim 1$ TeV.  

$R_D$, if it is the lightest fractionally charged state, is stable, as discussed in Sec.~\ref{subsec:ChargedRelics}; it will hadronize in the equivalent of a SUSY R-hadron, for which bounds require $m_{R} \gtrsim 1$ TeV~\cite{CMS:2016kce}. The field $\Delta_0$, if light enough, can also be pair-produced at colliders from two down-type quarks or single-produced with an $M$ from a gluon and a down-type quark, as the diagrams in the second row of Fig.~\ref{fig:PairProduction} show. The $\Delta_0$ lifetime is given in Eq.~\eqref{eq:Delta0lifetime}, implying a prompt decay $\Delta_0 \to 3 \text{ JET}s + \text{MET}$ at colliders. Squark searches constrain the mass of the stop to $m_{\tilde q} > 1.3$ TeV~\cite{ATLAS:2020syg} for such a channel. However, because the production cross-section for $\Delta_0$ occurs through down-quarks and the t-channel $M$ mediation, we expect it to be highly suppressed relative to pair-production of squarks via gluon fusion, weakening the constraint on $M_{\Delta_0}$. If the $\Delta_0$ lifetime is long enough to hadronize, their hybrids, which are singlets under the SM gauge group, could be traced by searching for monojet signals~\cite{ATLAS:2017bfj,ATLAS:2021kxv}.

The theory also predicts a light colored octet $\rho_8$ at the TeV scale. It can be pair-produced from gluons, and each of them decays as $\rho_8 \to 4 \text{ JET}s$ through a pair of $S_1$ off-shell. The $\rho_8$ lifetime depends on the parameters in the scalar potential, 
\begin{equation}
\tau_{\rho_8}^{-1} \lesssim \frac{1}{1024\pi^5} \left(\frac{|Y_\text{quirks}|}{1}\right)^4 \frac{M_{\rho_8}^9}{M_{S_1}^8} \simeq 5 \times 10^{13} \text{ s}^{-1}\left(\frac{|Y_\text{quirks}|}{1}\right)^4 \left(\frac{10 \text{ TeV}}{M_{S_1}}\right)^8 \left(\frac{M_{\rho_8}}{1 \text{ TeV}}\right)^9,
\end{equation}
where the symbol $\lesssim$ reflects that the coupling of $S_1^* \rho_8 S_1$ weighted by $v_{24}$ contributes to the $\rho_8$ mass. The decay of $\rho_8$ could lead to displaced vertices of four jets for smaller couplings than $M_{\rho_8}$. 

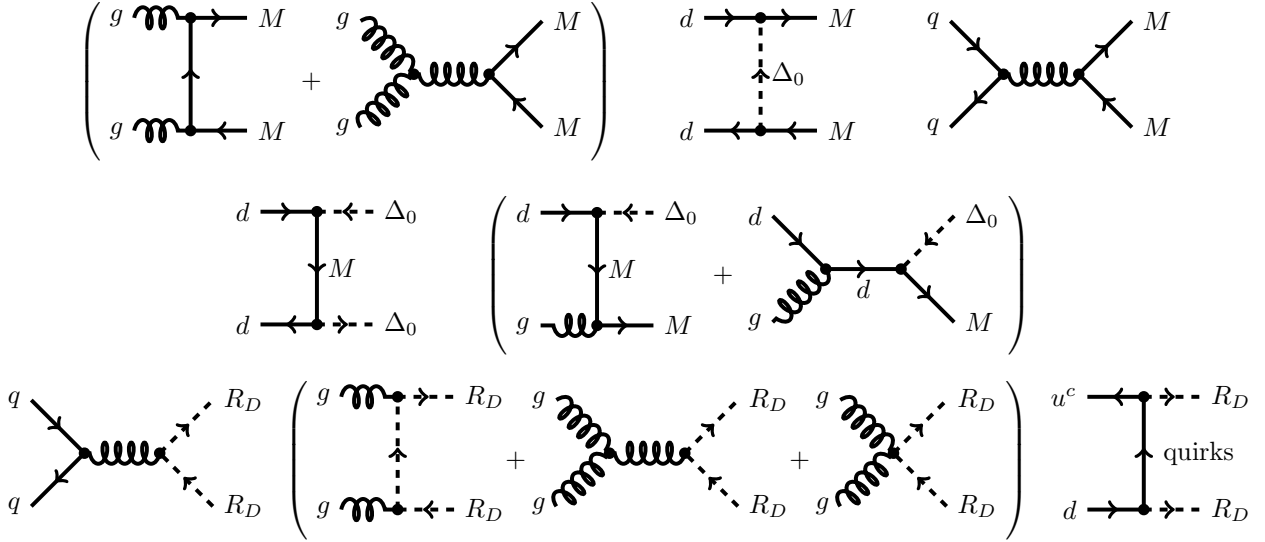
\begin{figure}[t]
\begin{equation*}
\left( 
\begin{gathered}
\begin{tikzpicture}[line width=1.5 pt,node distance=1 cm and 1.5 cm]
\coordinate[label=left:$g$](g1); 
\coordinate[right = 0.75 cm of g1](v1);
\coordinate[below =  1.5 cm of v1](v2);
\coordinate[left = 0.75 cm of v2,label=left:$g$](g2);
\coordinate[right = 0.75 cm of v1,label=right:$M$](M1);
\coordinate[right = 0.75 cm of v2,label=right:$M$](M2);

\draw[gluon](g1)--(v1);
\draw[gluon](g2)--(v2);
\draw[fermion](v1)--(M1);
\draw[fermion](M2)--(v2);
\draw[fermion](v2)--(v1);
\draw[fill=black] (v1) circle (.05cm);
\draw[fill=black] (v2) circle (.05cm);

\end{tikzpicture}
\end{gathered}
+ 
\begin{gathered}
\begin{tikzpicture}[line width=1.5 pt,node distance=1 cm and 1.5 cm]
\coordinate[label=left:$g$](g1); 
\coordinate[below right = 1 cm of g1](v1);
\coordinate[right =  1 cm of v1](v2);
\coordinate[below left = 1 cm of v1,label=left:$g$](g2);
\coordinate[above right = 1 cm of v2,label=right:$M$](M1);
\coordinate[below right = 1 cm of v2,label=right:$M$](M2);
\draw[gluon](g1)--(v1);
\draw[gluon](g2)--(v1);
\draw[fermion](v2)--(M1);
\draw[fermion](M2)--(v2);
\draw[gluon](v2)--(v1);
\draw[fill=black] (v1) circle (.05cm);
\draw[fill=black] (v2) circle (.05cm);
\end{tikzpicture}
\end{gathered} \right)
\qquad
\begin{gathered}
\begin{tikzpicture}[line width=1.5 pt,node distance=1 cm and 1.5 cm]
\coordinate[label=left:$d$](g1); 
\coordinate[right = 0.75 cm of g1](v1);
\coordinate[below =  1.5 cm of v1](v2);
\coordinate[left = 0.75 cm of v2,label=left:$ d$](g2);
\coordinate[right = 0.75 cm of v1,label=right:$M$](M1);
\coordinate[right = 0.75 cm of v2,label=right:$M$](M2);
\coordinate[below= 0.75 cm of v1,label=right:$\Delta_0$](aux);
\draw[fermion](g1)--(v1);
\draw[fermion](v2)--(g2);
\draw[fermion](v1)--(M1);
\draw[fermion](M2)--(v2);
\draw[scalar](v2)--(v1);
\draw[fill=black] (v1) circle (.05cm);
\draw[fill=black] (v2) circle (.05cm);

\end{tikzpicture}
\end{gathered}
\qquad 
\begin{gathered}
\begin{tikzpicture}[line width=1.5 pt,node distance=1 cm and 1.5 cm]
\coordinate[label=left:$q$](g1); 
\coordinate[below right = 1 cm of g1](v1);
\coordinate[right =  1 cm of v1](v2);
\coordinate[below left = 1 cm of v1,label=left:$ q$](g2);
\coordinate[above right = 1 cm of v2,label=right:$M$](M1);
\coordinate[below right = 1 cm of v2,label=right:$M$](M2);
\draw[fermion](g1)--(v1);
\draw[fermion](v1)--(g2);
\draw[fermion](v2)--(M1);
\draw[fermion](M2)--(v2);
\draw[gluon](v2)--(v1);
\draw[fill=black] (v1) circle (.05cm);
\draw[fill=black] (v2) circle (.05cm);
\end{tikzpicture}
\end{gathered} 
\end{equation*}
\begin{equation*}
\begin{gathered}
\begin{tikzpicture}[line width=1.5 pt,node distance=1 cm and 1.5 cm]
\coordinate[label=left:$d$](g1); 
\coordinate[right = 0.75 cm of g1](v1);
\coordinate[below =  1.5 cm of v1](v2);
\coordinate[left = 0.75 cm of v2,label=left:$d$](g2);
\coordinate[right = 0.75 cm of v1,label=right:$\Delta_0$](M1);
\coordinate[right = 0.75 cm of v2,label=right:$\Delta_0$](M2);
\coordinate[below= 0.75 cm of v1,label=right:$M$](aux);
\draw[fermion](g1)--(v1);
\draw[fermion](v2)--(g2);
\draw[scalar](M1)--(v1);
\draw[scalar](v2)--(M2);
\draw[fermion](v1)--(v2);
\draw[fill=black] (v1) circle (.05cm);
\draw[fill=black] (v2) circle (.05cm);
\end{tikzpicture}
\end{gathered}
\qquad 
\left( \begin{gathered}
\begin{tikzpicture}[line width=1.5 pt,node distance=1 cm and 1.5 cm]
\coordinate[label=left:$d$](g1); 
\coordinate[right = 0.75 cm of g1](v1);
\coordinate[below =  1.5 cm of v1](v2);
\coordinate[left = 0.75 cm of v2,label=left:$g$](g2);
\coordinate[right = 0.75 cm of v1,label=right:$\Delta_0$](M1);
\coordinate[right = 0.75 cm of v2,label=right:$M$](M2);
\coordinate[below= 0.75 cm of v1,label=right:$M$](aux);
\draw[fermion](g1)--(v1);
\draw[gluon](v2)--(g2);
\draw[scalar](M1)--(v1);
\draw[fermion](v2)--(M2);
\draw[fermion](v1)--(v2);
\draw[fill=black] (v1) circle (.05cm);
\draw[fill=black] (v2) circle (.05cm);
\end{tikzpicture}
\end{gathered}
+
\begin{gathered}
\begin{tikzpicture}[line width=1.5 pt,node distance=1 cm and 1.5 cm]
\coordinate[label=left:$d$](g1); 
\coordinate[below right = 1 cm of g1](v1);
\coordinate[right =  1 cm of v1](v2);
\coordinate[below left = 1 cm of v1,label=left:$g$](g2);
\coordinate[above right = 1 cm of v2,label=right:$\Delta_0$](M1);
\coordinate[below right = 1 cm of v2,label=right:$M$](M2);
\coordinate[right= 0.5 cm of v1,label=below:$d$](aux);
\draw[fermion](g1)--(v1);
\draw[gluon](v1)--(g2);
\draw[scalar](M1)--(v2);
\draw[fermion](v2)--(M2);
\draw[fermion](v1)--(v2);
\draw[fill=black] (v1) circle (.05cm);
\draw[fill=black] (v2) circle (.05cm);
\end{tikzpicture}
\end{gathered} \right)
\end{equation*}
\begin{equation*}
\begin{gathered}
\begin{tikzpicture}[line width=1.5 pt,node distance=1 cm and 1.5 cm]
\coordinate[label=left:$q$](g1); 
\coordinate[below right = 1 cm of g1](v1);
\coordinate[right =  1 cm of v1](v2);
\coordinate[below left = 1 cm of v1,label=left:$q$](g2);
\coordinate[above right = 1 cm of v2,label=right:$R_D$](M1);
\coordinate[below right = 1 cm of v2,label=right:$R_D$](M2);
\draw[fermion](g1)--(v1);
\draw[fermion](v1)--(g2);
\draw[scalar](v2)--(M1);
\draw[scalar](M2)--(v2);
\draw[gluon](v2)--(v1);
\draw[fill=black] (v1) circle (.05cm);
\draw[fill=black] (v2) circle (.05cm);
\end{tikzpicture}
\end{gathered}
\
\left(
\begin{gathered}
\begin{tikzpicture}[line width=1.5 pt,node distance=1 cm and 1.5 cm]
\coordinate[label=left:$g$](g1); 
\coordinate[right = 0.75 cm of g1](v1);
\coordinate[below =  1.5 cm of v1](v2);
\coordinate[left = 0.75 cm of v2,label=left:$g$](g2);
\coordinate[right = 0.75 cm of v1,label=right:$R_D$](M1);
\coordinate[right = 0.75 cm of v2,label=right:$R_D$](M2);
\draw[gluon](g1)--(v1);
\draw[gluon](g2)--(v2);
\draw[scalar](v1)--(M1);
\draw[scalar](M2)--(v2);
\draw[scalar](v2)--(v1);
\draw[fill=black] (v1) circle (.05cm);
\draw[fill=black] (v2) circle (.05cm);
\end{tikzpicture}
\end{gathered} 
\!\!\! + \!\!\!   
\begin{gathered}
\begin{tikzpicture}[line width=1.5 pt,node distance=1 cm and 1.5 cm]
\coordinate[label=left:$g$](g1); 
\coordinate[below right = 1 cm of g1](v1);
\coordinate[right =  1 cm of v1](v2);
\coordinate[below left = 1 cm of v1,label=left:$g$](g2);
\coordinate[above right = 1 cm of v2,label=right:$R_D$](M1);
\coordinate[below right = 1 cm of v2,label=right:$R_D$](M2);
\draw[gluon](g1)--(v1);
\draw[gluon](g2)--(v1);
\draw[scalar](v2)--(M1);
\draw[scalar](M2)--(v2);
\draw[gluon](v2)--(v1);
\draw[fill=black] (v1) circle (.05cm);
\draw[fill=black] (v2) circle (.05cm);
\end{tikzpicture}
\end{gathered} 
\!\!\! + \!\!\!   
\begin{gathered}
\begin{tikzpicture}[line width=1.5 pt,node distance=1 cm and 1.5 cm]
\coordinate[label=left:$g$](g1); 
\coordinate[below right = 1 cm of g1](v1);
\coordinate[below left = 1cm of v1,label=left:$g$](g2);
\coordinate[above right = 1 cm of v1,label=right:$R_D$](M1);
\coordinate[below right = 1 cm of v1,label=right:$R_D$](M2);
\draw[gluon](g1)--(v1);
\draw[gluon](g2)--(v1);
\draw[scalar](v1)--(M1);
\draw[scalar](M2)--(v1);
\draw[fill=black] (v1) circle (.05cm);
\end{tikzpicture}
\end{gathered}
\right)
\
\begin{gathered}
\begin{tikzpicture}[line width=1.5 pt,node distance=1 cm and 1.5 cm]
\coordinate[label=left:$u^c$](g1); 
\coordinate[right = 0.75 cm of g1](v1);
\coordinate[below =  1.5 cm of v1](v2);
\coordinate[left = 0.75 cm of v2,label=left:$d$](g2);
\coordinate[right = 0.75 cm of v1,label=right:$R_D$](M1);
\coordinate[right = 0.75 cm of v2,label=right:$R_D$](M2);
\coordinate[below= 0.75 cm of v1,label=right:$\text{ quirks}$](aux);
\draw[fermion](v1)--(g1);
\draw[fermion](g2)--(v2);
\draw[scalar](v1)--(M1);
\draw[scalar](v2)--(M2);
\draw[fermion](v2)--(v1);
\draw[fill=black] (v1) circle (.05cm);
\draw[fill=black] (v2) circle (.05cm);
\end{tikzpicture}
\end{gathered}  
\end{equation*}
\caption{Feynman diagrams for the pair-production of potentially light connector fields. The pair-production of the diquark $S_1$ involves the same diagrams as $R_D$ except the last one, which should be substituted by the t-channel process $\bar d \, d  \, (\text{or }\bar u \, u ) \to S_1^* S_1$.}
\label{fig:PairProduction}
\end{figure}

\item {\bf Flavor.}  With only one flavor, the dark matter itself does not have a flavor structure.  But theories of Asymmetric Dark Matter are inherently flavorful (as discussed in Ref.~\cite{Kim:2013ivd}), and the mediator states appearing in the UV completion of the ADM operators themselves mediate flavor signatures.   

The connectors $M$, $S_1$ and $\Delta_0$ participating in the darko-baryo-genesis could be at the TeV scale, motivated by cosmology. They interact with the SM quarks through the (vector) Yukawa coupling $Y_{\Delta_0}$ (only one chirality available) or the Yukawa matrix $Y_\text{quirks}$ (both chiralities present). We expect the strongest bounds on these interactions coming from the constraints on neutral meson mixing, generated by the following box diagrams:
\begin{equation}
\begin{gathered}
\begin{tikzpicture}[line width=1.5 pt,node distance=1 cm and 1.5 cm]
\coordinate[label=left:$d_i$](di);
\coordinate[right = 0.75cm of di](v1);
\coordinate[right = of v1](v3);
\coordinate[right = 0.75 cm  of v3,label=right:$d_j$](df);
\coordinate[below = 1.5 cm of v1](v2);
\coordinate[left = 0.75 cm of v2,label=left:$ d_j$](si);
\coordinate[below = 1.5 cm of v3](v4);
\coordinate[right = 0.75 cm of v4,label=right:$ d_i$](sf);
\coordinate[below = 0.75 cm of v1,label=left:$S_1$](auxW);
\coordinate[below = 0.75 cm of v3, label=right:$S_1$](auxE);
\coordinate[right = 0.75 cm of v1, label=above:$u^c$](auxN);
\coordinate[right = 0.75 cm of v2,label=below:$u^c$](auxS);
\draw[fermion](di)--(v1);
\draw[scalar](v1)--(v2);
\draw[fermion](v2)--(si);
\draw[fermion](sf)--(v4);
\draw[scalar](v4)--(v3);
\draw[fermion](v3)--(df);
\draw[fermion](v1)--(v3);
\draw[fermion](v4)--(v2);
\draw[fill=black] (v1) circle (.05cm);
\draw[fill=black] (v2) circle (.05cm);
\draw[fill=black] (v3) circle (.05cm);
\draw[fill=black] (v4) circle (.05cm);
\end{tikzpicture}
\end{gathered}  \! , 
\qquad 
\begin{gathered}
\begin{tikzpicture}[line width=1.5 pt,node distance=1 cm and 1.5 cm]
\coordinate[label=left:$d_i$](di);
\coordinate[right = 0.75cm of di](v1);
\coordinate[right = of v1](v3);
\coordinate[right = 0.75 cm  of v3,label=right:$d_j$](df);
\coordinate[below = 1.5 cm of v1](v2);
\coordinate[left = 0.75 cm of v2,label=left:$ d_j$](si);
\coordinate[below = 1.5 cm of v3](v4);
\coordinate[right = 0.75 cm of v4,label=right:$ d_i$](sf);
\coordinate[below = 0.75 cm of v1,label=left:$M$](auxW);
\coordinate[below = 0.75 cm of v3, label=right:$M$](auxE);
\coordinate[right = 0.75 cm of v1, label=above:$\Delta_0$](auxN);
\coordinate[right = 0.75 cm of v2,label=below:$\Delta_0$](auxS);
\draw[fermion](di)--(v1);
\draw[fermion](v1)--(v2);
\draw[fermion](v2)--(si);
\draw[fermion](sf)--(v4);
\draw[fermion](v4)--(v3);
\draw[fermion](v3)--(df);
\draw[scalar](v3)--(v1);
\draw[scalar](v2)--(v4);
\draw[fill=black] (v1) circle (.05cm);
\draw[fill=black] (v2) circle (.05cm);
\draw[fill=black] (v3) circle (.05cm);
\draw[fill=black] (v4) circle (.05cm);
\end{tikzpicture}
\end{gathered}.
\label{eq:boxdiagrams}
\end{equation}
The left-hand diagram is a standard diquark process, and the bounds are computed in Ref.~\cite{Giudice:2011ak}
The right-hand diagram is similar, but has dark color running in the loop and a different chiral structure.  We find a bound
\begin{equation}
\sqrt{\left|\text{Re}  \{   (Y_{\Delta_0}^1 (Y_{\Delta_0}^2)^* )^2  \} \right |} \left(\frac{\text{TeV}}{M_{M/\Delta_0}}\right)\lesssim 0.01, \ \ \sqrt{\left|\text{Im}  \{   (Y_{\Delta_0}^1 (Y_{\Delta_0}^2)^* )^2  \} \right |}\left(\frac{\text{TeV}}{M_{M/\Delta_0}}\right)\lesssim 8 \times 10^{-4},
\label{eq:boundMMbar}
\end{equation}
derived from saturating the experimental measurement on the kaon mass difference $\Delta m_K = (3.484 \pm 0.006) \times 10^{-15} \text{ GeV}$ and $|\epsilon_K| = (2.228 \pm 0.011) \times 10^{-3}$~\cite{Zyla:2020zbs}.  

The $M$ and $\Delta_0$ do not contribute to the electric dipole moment (EDM) of the quarks at leading order, but the diquark $S_1$ does
\begin{equation}
|d_d| \simeq \sum_j \frac{e}{6\pi^2}\frac{m_{q_j}}{M_{S_1}^2}\text{Im} \{Y_\eta^{1j} (Y_\Psi^{1j})^* \} \ln \left (\frac{M_{S_1}}{m_{q_j}}\right),
\end{equation}
where $m_{q_j}$ is the mass of the quark running inside the loop. If the quark is a top, the neutron EDM $d_n^\text{exp} \leq 2.9 \times 10^{-26} \ e \, \text{cm}$~\cite{Baker:2006ts} imposes 
\begin{equation}
\frac{\text{Im} \{Y_\text{quirks} \} }{M_{S_1}/\text{TeV}} \lesssim 4 \times 10^{-4}.
\end{equation}
The above constraints from neutral meson mixing and quark EDMs can be relaxed by assuming a certain flavor structure for the Yukawa vector. 

If the $R_D$ scalar is the lightest mixed field, the leading contributions to flavor observables are suppressed by the {\it quirk} masses $\sim v_{10}$, and are therefore consistent with the existing flavor constraints even for order one Yukawa couplings. 

\end{itemize}

\subsection{Astrophysical}

\begin{itemize}

\item {\bf Dark matter self-interactions.}  Because of the relatively low confinement scale of the dark baryons, we expect there to be a quite large dark matter self-interaction rate, set by the confinement scale:
\beq
\sigma/m_\rho \sim 16 \pi/m_\rho^3 \lesssim 1 \mbox{ cm}^2/\mbox{g},
\eeq
where the bound is heuristically quoted from self-interaction constraints~\cite{Tulin:2013teo}. This corresponds to a limit $m_\rho \gtrsim 500 \mbox{ MeV}$, which is easily met in this model.  On the other hand, we know that, where strong couplings are involved, the scattering cross-section can be resonantly enhanced by the presence of di-baryon bound states (the dark analogue of the deuteron).  Whether such a state exists would require lattice simulations, but would be the first step in the synthesis of larger bound states, called nuggets.  

\item {\bf Nugget formation.}  Similar to standard model nuclei, larger $N$ bound states could, in principle, be formed from the $\chi$ baryons.  Because the $\chi$ baryons are spin-1 states, they have no degeneracy pressure, and we expect the $\chi$ baryons to melt into their fundamental constituents as they form into larger nuggets.  As shown in previous studies~\cite{Gresham:2017cvl}, because of the absence of dark electromagnetism, there is no Coulomb barrier to the formation of larger states, and extremely large dark nuclei naturally form. The barrier to synthesizing such larger states is whether there are lower dark-number bound states (the dark analogue of the deuteron) that form as a gateway to larger states, as studied for fermionic asymmetric dark matter in Ref.~\cite{Wise:2014jva}.  Determining the existence of such states, is beyond the scope of this work, but could impact in important qualitative ways the astrophysics and detection of the dark matter.

\item {\bf Gravitational Waves.}  The dark phase transition, from a Hidden Valley, at temperatures around 10 GeV could generate gravitational waves at a frequency of $10^{-9} - 10^{-8} \mbox{ Hz}$, if the transition is strongly first order.    This is in the range of future pulsar timing arrays such as SKA~\cite{Schwaller:2015tja}, and can be constrained or observed (to a lesser degree) with current arrays \cite{NANOGrav:2021flc}.  However, it is not known whether an $\SU(2)$ gauge group with 1 light flavor would generate a strong first order phase transition; this would be an interesting future study.   In addition, we have the spontaneous breaking of two $\text{U}(1)$'s to hypercharge and the accidental $\text{U}(1)_Q$ to a ${\cal Z}_2$, which generates cosmic strings that in turn radiate energy through gravitational waves. The latter are washed out if inflation occurs at the low scale.

\item {\bf Early Matter Domination.}   As Eq.~\eqref{eq:Delta0lifetime} and~\eqref{eq:lifetimeN1} show, the connector $\Delta_0$ and the sterile neutrinos $N$, which enter in the baryo-darko-genesis in this model, as well as the dark hadrons (glueballs, $\eta$, see Eqs.~\eqref{eq:etalifetime},~\eqref{eq:GammaGB}) can be quite long lived,   decaying at temperatures well below its mass.  This implies a period of Early Matter Domination, where metric perturbations on small scales can enter the horizon and grow.  Scales that were inside the horizon during this period of EMD, can give rise to enhanced matter power spectrum on small scales, corresponding to subhalos of mass $10^{-15}-10^{-11} \, M_\odot$ for $T_\text{RH} = 1$ GeV and $10^{-9}-10^{-4} \, M_\odot$ for $T_\text{RH} = 1 \text{ MeV}$~\cite{Lee:2020wfn}.  Such subhalos may eventually be measured through, {\it e.g.}, pulsar timing arrays or photometric measurements, another evidence for low scale baryo-darko-genesis.

\end{itemize}

\section{Potential embedding in $\SU(7)$}
\label{sec:SU7}
It is very attractive to think of embedding the simplest non-abelian extension of the SM connecting the QCD and dark confinement scales within a Grand Unified Theory (GUT) framework. In this case, the minimal GUT where $\SU(5) \otimes \SU(2)_L \otimes \text{U}(1)_X$ can be embedded is $\SU(7)$. The first sign for motivation towards Grand Unification appears when realizing that one generation of SM fermions and a potential dark matter candidate are embedded in the lowest dimensional representations of $\SU(7)$: the anti-fundamental $\bar 7$ and the anti-symmetric $21$,
\begin{equation}
    \bar 7 = \begin{pmatrix} d^c  \\ \chi \\ -- \\ \ell \end{pmatrix}, \qquad \text{ and } \qquad 21 = \begin{pmatrix} u^c & M & | & q \\ - M^T & N & | & \Psi_2 \\ 
    ---&--- & | & -- \\
    -q^T & -\Psi_2^T & | & e^c
    \end{pmatrix}.
\end{equation}
Along with the SM fields and the DM $\chi \sim (1,1,0,2)$, a sterile neutrino $N \sim (1,1,0,1)$, plus messengers $M \sim (3,1,-1/3,2)$ and $\Psi_2 \sim (1,2,1/2,2)$ are predicted. We expect the existence of other fermion representations to render the theory anomaly free.

The symmetry breaking pattern of $\SU(5)$ can be quite naturally achieved within the confines of the larger $\SU(7)$ structure. The theory requires the presence of the adjoint $48_H$ in order to break $\SU(7)$ without lowering the rank:
\begin{equation}
    \SU(7) \stackrel{\langle 48_H \rangle}{\to } \SU(5) \otimes \SU(2)_L \otimes \text{U}(1)_7
\end{equation}
where $\text{U}(1)_7$ can be identified with the previous $\text{U}(1)_X$. Note that, up to two a global normalization, all $\text{U}(1)_7$ hypercharges are fixed by the breaking of $\SU(7)$. Inside the $48_H$, we have already the $24_H$ representation that will trigger the breaking of $\SU(5)$ in the desired way, 
\begin{equation}
    \SU(5) \otimes \SU(2)_L \otimes \text{U}(1)_7 \stackrel{\langle 24_H \rangle}{\to} \SU(3)_C \otimes \SU(2)_L \otimes \text{U}(1)_7 \otimes \text{U}(1)_5 \otimes \SU(2)_D.
\end{equation}
Finally, as argued before, we need a singlet under the non-abelian groups, charged under $\text{U}(1)_7$ and $\text{U}(1)_5$ with zero hypercharge, such that its vev breaks the two $\text{U}(1)$ to the diagonal $\text{U}(1)_Y$. For such breaking we already know that the antisymmetric of $\SU(5)$ is required, which lives in the antisymmetric of $\SU(7)$, {\it i.e.} the $\delta_0 \sim (1,1,0,1) \subset 21_H$,
\begin{equation}
\text{U}(1)_7 \otimes \text{U}(1)_5 \stackrel{\langle 21_H \rangle}{\to} \text{U}(1)_Y.
\end{equation}

The generation of the fermion masses gives a larger structure that contains in it many of the interactions highlighted before. In order to recover the SM structure, we consider embedding the SM Higgs in the fundamental of $\SU(7)$:
\begin{equation}
   7_H \sim \begin{pmatrix} S_1^* \\ \Delta_0 \\--\\ H \end{pmatrix},
\end{equation}
where we use the same notation as in previous sections, {\it i.e.} $S_1 \sim (\bar 3,1,1/3,1)$ and $\Delta_0 \sim (1,1,0,2)$.  We generate the Yukawa term
\begin{equation}
  Y_e \,  \bar 7 \, 21 \, 7_H^*. 
\end{equation}
If we expand it over the SM components, it gives rise to the following interactions:
\begin{equation}
\begin{split}
    Y_e \left ( d^c  u^c S_1 + d^c  M  \Delta_0^* + \chi  M  S_1 +  \chi  N  \Delta_0^* + d^c  q  H^\dagger + \chi  \Psi_2  H^\dagger + \ell  \, q  \,  S_1 + \ell \, \Psi_2 \Delta_0^* + \ell e^c H^\dagger \right),
    \end{split}
\end{equation}
which contain those needed for darko-baryo-genesis in the ${\cal G}_5$ theory.  As expected, there are now additional interactions simultaneously with quarks and leptons that imply the mass of $S_1$ to be at the GUT scale.  The interaction $\chi \, \Psi_2 H^\dagger$ mixes the dark matter $\chi$ with the $\SU(2)$ neutral component of $\Psi_2$ through an EW mass term.   In order to make this effective mixing small, $\Psi_2$ must have a Dirac mass term that is significantly heavier (TeV scale or higher) than the $\chi$ mass.  

We have highlighted here the main features of the $\SU(7)$ completion that naturally generate the $\SU(5)$ structure of Dark Unification and the dynamics of darko-baryo-genesis.  This includes notably the main characters of the cast: the singlet fermion $N$, and its interactions with the scalar and fermion states of $\SU(5)$, particularly the connectors $M$, $\Delta_0$ and $S_1$, and the dark matter $\chi$.  It also includes the needed scalars for consistent breaking of $\SU(7)$ to $\SU(5)$ and finally the SM plus the dark $\SU(2)$, {\it i.e.} the supporting characters.  Additional interactions will be required to fill out the entire fermionic mass structure of the Unified $\SU(7)$, but we leave this to future work, having focused here on Dark Unification.

\section{Conclusions}

We have proposed a Dark-Color Unification theory, where the dark matter to baryon mass ratios are fixed by the ratio of the dark and visible color confinement scales.  The unification of the two sectors at a high scale guarantees that these confinement scales are nearby.  By simultaneously fixing the dark matter and baryon number densities through darko-baryo-genesis, the observed ratio dark matter to baryon mass density ratio, $\rho_D/\rho_B$, is easily derived.

The rich structure of the Unification sector provides many features that both provide for a cosmology consistent with observation, as well as the states that mediate darko-baryo-genesis.  Most notably, these consist of messenger states that mediate the lightest dark hadron decay, as well as scalars that transfer the dark matter and baryon asymmetries to the respective sectors.  

There is a plethora of possible signatures from such a model, that demand further study.  Because of the low confinement scale in the hidden sector, we expect large dark matter self-interactions that could be observed.  There is typically early matter domination from the late decay of the unstable dark hadrons,
as well as gravitational waves should the dark sector have a strong first order dark color phase transition.  There are also possible flavor signals, generated by connector states that can also potentially be pair produced at colliders, decaying to standard model jets plus dark sector states, with or without displaced vertices.
We look forward to exploring some of these signatures in future work.

\begin{center}
{\bf Acknowledgments }\\
\end{center}
We thank Andrea Mitridate, Michele Papucci and Mario Reig for discussions, as well as Pavel Fileviez-Perez and Mark B. Wise for comments on the draft.  The work of KZ is supported by the DoE under contract DE-SC0011632, and by a Simons Investigator award. This work is also supported by the Walter Burke Institute for Theoretical Physics.

\appendix

\bibliography{DarkoBib.bib}

\end{document}